\documentclass[lettersize,journal]{IEEEtran}
\usepackage{amsmath,amsfonts}
\usepackage{algorithmic}
\usepackage{array}
\usepackage{textcomp}
\usepackage{stfloats}
\usepackage{url}
\usepackage{verbatim}
\usepackage{graphicx}
\def\BibTeX{{\rm B\kern-.05em{\sc i\kern-.025em b}\kern-.08em
    T\kern-.1667em\lower.7ex\hbox{E}\kern-.125emX}}
\usepackage{balance}
\usepackage{mdframed}
\usepackage{svg}
\usepackage{booktabs}

\usepackage{overpic}
\usepackage{xspace}
\usepackage{enumitem}
\usepackage{multicol}

\usepackage{color}
\usepackage{colortbl}
\usepackage{soul}
\usepackage{subcaption}
\usepackage{fancyhdr} 
\usepackage{float}
\usepackage{multirow}
\usepackage{wrapfig}
\usepackage{makecell}
\usepackage{tabularx}
\usepackage{refcount}
\usepackage{tcolorbox}
\usepackage{soul}
\usepackage[misc]{ifsym}
\usepackage{ifthen}
\usepackage{hyperref}
\newboolean{doublecolumn}
\setboolean{doublecolumn}{true} 

\tcbuselibrary{listings}

\newtcolorbox[auto counter]{tcbfinding}[1][]{
    top=3pt,
    bottom=3pt,
    #1
}

\usepackage{scalerel}
\usepackage{tikz}
\usetikzlibrary{svg.path}
\definecolor{orcidlogocol}{HTML}{A6CE39}
\tikzset{
  orcidlogo/.pic={
    \fill[orcidlogocol] svg{M256,128c0,70.7-57.3,128-128,128C57.3,256,0,198.7,0,128C0,57.3,57.3,0,128,0C198.7,0,256,57.3,256,128z};
    \fill[white] svg{M86.3,186.2H70.9V79.1h15.4v48.4V186.2z}
                 svg{M108.9,79.1h41.6c39.6,0,57,28.3,57,53.6c0,27.5-21.5,53.6-56.8,53.6h-41.8V79.1z M124.3,172.4h24.5c34.9,0,42.9-26.5,42.9-39.7c0-21.5-13.7-39.7-43.7-39.7h-23.7V172.4z}
                 svg{M88.7,56.8c0,5.5-4.5,10.1-10.1,10.1c-5.6,0-10.1-4.6-10.1-10.1c0-5.6,4.5-10.1,10.1-10.1C84.2,46.7,88.7,51.3,88.7,56.8z};
  }
}
\newcommand\orcidicon[1]{\href{https://orcid.org/#1}{\mbox{\scalerel*{
\begin{tikzpicture}[yscale=-1,transform shape]
\pic{orcidlogo};
\end{tikzpicture}
}{|}}}}

\makeatletter
\DeclareRobustCommand\onedot{\futurelet\@let@token\@onedot}
\def\@onedot{\ifx\@let@token.\else.\null\fi\xspace}
\def\eg{{e.g}\onedot} 
\def\ie{{i.e}\onedot} 
 
\def\etc{{etc}\onedot} 
 
\def\etal{{et al}\onedot}
\makeatother

\definecolor{lightgray}{HTML}{eeeeee}

\definecolor{highlightColor}{rgb}{1, 0.8, 0.6}

\definecolor{amii_magenta}{HTML}{bf477c}
\definecolor{amii_summer}{HTML}{ffcccc}
\definecolor{amii_mustard}{HTML}{faa53c}
\definecolor{amii_sky}{HTML}{6c98ab}
\definecolor{amii_emerald}{HTML}{006c65}
\definecolor{amii_night}{HTML}{003f58}

\definecolor{top1Color}{HTML}{E7AFB6}
\definecolor{top2Color}{HTML}{ffb74d}
\definecolor{top3Color}{HTML}{ffecb3}
\definecolor{jiayang_todo}{HTML}{2FAF38}
\sethlcolor{highlightColor}

\newcommand{\toponehl}[1]{\sethlcolor{top1Color}\hl{#1}}
\newcommand{\toptwohl}[1]{\sethlcolor{top2Color}\hl{#1}}
\newcommand{\topthreehl}[1]{\sethlcolor{top3Color}\hl{#1}}

\newtcolorbox{questionbox}{%
    colback=blue!5!white,
    colframe=blue!75!black,
    title=Question:
}

\newtcolorbox{answerbox}{%
    colback=green!5!white,
    colframe=green!75!black,
    title=Answer:
}

\newcommand{\responseref}[0]{\color{black}}
\newcommand{\responseline}[1]{\textcolor{black}{#1}}

\newcommand{\jiayang}[1]{\textcolor{jiayang_todo}{(JS) #1}}

\newcommand{\rqone}{To what extent can the uncertainty estimation techniques help identify potential risks of LLMs in NLP tasks?}
    
\newcommand{\rqtwo}{What limitations do the uncertainty estimation methods encounter when applied to LLMs in the context of NLP tasks?}

\newcommand{\rqthree}{To what extent can the uncertainty estimation methods assist in identifying potential risks of LLMs for code generation?}

\newcommand{\rqfour}{What potential limitation do the uncertainty estimation methods face when being applied to LLMs for code generation?}

\newif\ifdisplaycontent
\displaycontenttrue  

\begin{document}
\title{
\textit{Look Before You Leap:}
{\responseline{An Exploratory Study of Uncertainty Analysis for Large Language Models}}}


\author{
        Yuheng Huang, 
        Jiayang Song, 
        Zhijie Wang, 
        Shengming Zhao, 
        Huaming Chen, 
        Felix~Juefei-Xu, 
        and Lei Ma$^{\textrm{\Letter}}$
        \thanks{Yuheng Huang is is with The University of Tokyo, Tokyo 113-8658, Japan (e-mail: yuhenghuang42@g.ecc.u-tokyo.ac.jp).}
        \thanks{Jiayang Song, Zhijie Wang and Shengming Zhao are with the University of Alberta, Edmonton, T6G 1H9, Canada (e-mail: \{jiayan13, zhijie.wang, shengmi1\}@ualberta.ca).}
        \thanks{Huaming Chen is with The University of Sydney, Australia (e-mail: huaming.chen@sydney.edu.au).}
        \thanks{Felix Juefei-Xu is with New York University, New York, NY 10012, USA (e-mail: juefei.xu@nyu.edu).}
        \thanks{Lei Ma is with The University of Tokyo, Tokyo 113-8658, Japan, and also with the University of Alberta, Canada (e-mail: ma.lei@acm.org).}
        \thanks{$^{\textrm{\Letter}}$ Lei Ma is the corresponding author.}
}

\IEEEtitleabstractindextext{%
    \begin{abstract}
        
        The recent performance leap of Large Language Models (LLMs) opens up new opportunities across numerous industrial applications and domains.
        However, the potential erroneous behavior (e.g., the generation of misinformation and hallucination) has also raised severe concerns for the trustworthiness of LLMs, especially in safety-, security- and reliability-sensitive industrial scenarios, potentially hindering real-world adoptions.
        While uncertainty estimation has shown its potential for interpreting the prediction risks made by classic machine learning (ML) models, the unique characteristics of recent LLMs (e.g., adopting self-attention mechanism as its core, very large-scale model size, often used in generative contexts) 
        pose new challenges for the behavior analysis of LLMs. 
        Up to the present, little progress has been made to better understand whether and to what extent uncertainty estimation can help characterize the capability boundary of an LLM, to counteract its undesired behavior, which is considered to be of great importance with the potential wide-range applications of LLMs across industry domains.
        {
        To bridge the gap, in this paper, we initiate an early exploratory study of the risk assessment of LLMs from the lens of uncertainty. 
        In particular, we conduct a large-scale study with as many as twelve uncertainty estimation methods and \responseline{eight general LLMs on four NLP tasks and seven programming-capable LLMs on two code generation tasks} to investigate to what extent uncertainty estimation techniques could help characterize the prediction risks of LLMs. 
        Our findings confirm the potential of uncertainty estimation for revealing LLMs' uncertain/non-factual predictions. The insights derived from our study can pave the way for more advanced analysis and research on LLMs, ultimately aiming at enhancing their trustworthiness.
    }
    \end{abstract}
    
}

\markboth{IEEE Transactions on Software Engineering,~Vol.~, No.~, ~2024}%
{}

\maketitle
\IEEEdisplaynontitleabstractindextext

%
\IEEEpeerreviewmaketitle

\section{Introduction}
\label{sec:introduction}
\IEEEPARstart{L}{arge} Language Models (LLMs) have demonstrated impressive capabilities in miscellaneous Natural Language Processing (NLP) tasks and promising adaptability in practical applications across diverse domains, including but not limited to content moderation~\cite{markov2023holistic}, code generation~\cite{roziere2023code}, conversational AI~\cite{chatgpt2023}, and personalized content recommendations~\cite{wu2023survey}. The scale of deployment is vast, addressing the needs of diverse user demographics and industries. As a prominent example, Meta has launched foundation models such as the Llama family~\cite{touvron2023llama, touvron2023llama2}. By September 2023, these models had driven the creation of over 3,500 enterprise projects and inspired more than 7,000 GitHub repositories~~\cite{meta2023llamaecosystem}.


\begin{figure}[tbp]
     \centering
     \includegraphics[width=0.99\linewidth]{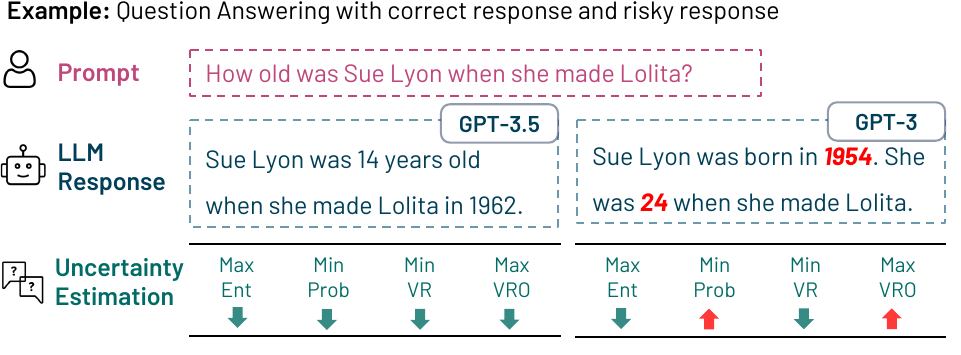}
     \caption{Uncertainty estimation for a QA task.}
     \vspace{-10pt}
     \label{fig:bg:example_QA}
\end{figure}

\begin{figure*}[t]
    \centering
    \includegraphics[width=0.85\linewidth]{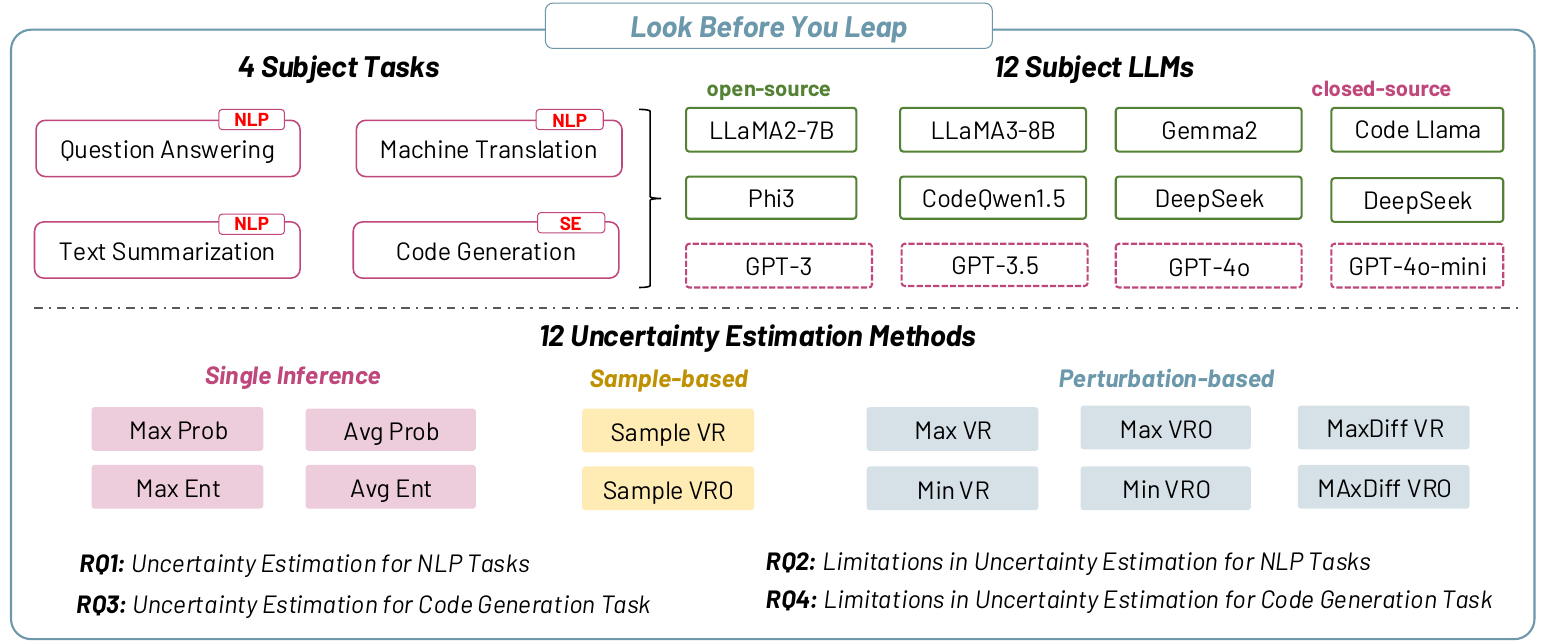}
    \caption{A high-level overview and workflow of this paper.}
    \label{fig:overall_workflow}
\end{figure*}






Despite the attractive performance that LLMs present and their rapid evolution within both academics and industries, an urgent common concern about LLMs has the propensity of generating erroneous information without warning.   
Such phenomenon of erroneous generation can exhibit in terms of different manifestations (\eg,  hallucination \cite{ji2023survey}, disinformation \cite{tamkin2021understanding}, bias \cite{abid2021persistent}) across various tasks.
In general, the current LLMs are found to have the tendency to generate problematic, nonfactual responses that are not from training sources or misguided by biased data. However, these responses are often presented in a natural human-like tone \cite{bang2023multitask, johnson2023assessing}. 
Such characteristics cause erroneous information to be highly mixed and intertwined with confident and factual contexts, making their detection and localization difficult without close inspection and diligent fact-checking \cite{bubeck2023sparks}.
As an example, Fig.~\ref{fig:bg:example_QA} (GPT-3) depicts an example of an LLM answering a question with nonfactual information.



Risk assessments thus become crucial in the process of mitigating such threats. A recent survey highlights that 98\% of respondents, encompassing domain experts and civil society members, firmly believe that AGI (artificial general intelligence) labs should undertake risk assessments before deployment~\cite{schuett2023towards}. For the AI industry, implementing comprehensive risk assessment methods is not just a technical necessity but also an ethical obligation. Major tech corporations~\cite{anderljung2023frontier} such as Microsoft~\cite{microsoft2022responsibleai}, OpenAI~\cite{openai2023aisafety}, Amazon~\cite{amazon2023responsibleai}, and Google~\cite{google2023responsibleai}, along with non-governmental organizations (NGOs, \eg, the Centre for the Governance of AI~\cite{koessler2023risk}), are fervently working towards developing safe, secure, transparent, reliable and responsible LLMs and AGI applications. As a driving force behind open and collaborative AI research, Meta also commits substantial resources to the development of responsible AI, emphasizing trustworthiness, transparency, robustness, etc.~\cite{meta2021fivepillars, meta2023responsibleai}. These endeavors encompass a range of reports (\eg, \textit{Building Generative AI Responsibly~\cite{meta2023generativeai}}), open-source tools (\eg, the model interpretability framework \textit{Captum}~\cite{kokhlikyan2020captum}), and datasets (\eg, \textit{Hateful Memes}~\cite{kiela2021hateful}).


Although there has been substantial work on other AI models, risk assessments for LLMs are still in their infancy. Due to their billions of parameters, vast amounts of (often inaccessible) data, and potential closed-source nature, LLMs present significant challenges for analysis and safeguarding. Yet, such a safeguard is crucial, especially considering the widespread adoption of LLMs. Stakeholders from research communities, industry, open source initiatives, NGOs, and businesses may all be negatively affected by untrustworthy LLMs. Uncertainty estimation, aimed at gauging the confidence level of model outputs~\cite{bhatt2021uncertainty, hullermeier2021aleatoric, rahmati2019predicting}, stands out as a promising approach for identifying risks in general Machine Learning (ML) models. Such techniques also have the potential for detecting erroneous generation from LLMs~\cite{manakul2023selfcheckgpt} even under black-box settings. {It is thus possible to take them as plug-and-play tools in both academic and industrial scenarios.} For example, Fig.~\ref{fig:bg:example_QA} shows that a higher uncertainty score could possibly indicate an erroneous generation of an LLM. 
{However, it is still unclear whether and to what extent uncertainty estimation methods could do when measuring and characterizing an LLM's capability limitations. Furthermore, it also raises questions such as \emph{``Are there better practices for employing these methods in practical scenarios''},} 
{\emph{``Do we need further adaptations from the industrial perspective to better cater to the distinct features of LLMs (\eg, task diversity and high computational cost)''}}, \etc.
To the best of our knowledge, up to the present, there is a lack of a general framework that integrates different uncertainty estimation methods for LLMs, as well as a systematic study to investigate the effectiveness of uncertainty estimation in characterizing an LLMs' capabilities. 

To bridge this gap, in this paper, we present an exploratory study to understand the trustworthiness of LLMs from the lens of uncertainty estimation.
Considering the generality and versatility for various application scenarios, we strive to identify suitable methods to minimize the requirement of LLMs internal information (\eg, model architecture, model parameters). 
{Such criteria enable these methods to be seamlessly adapted and incorporated by end-users of commercial models, such as GPT-3.5 and GPT-4}. Overall, we collected and implemented as many as 12 representative uncertainty estimation methods that were originally designed for general DNNs and successfully adapted them to the contexts of LLM applications.
To better capture an in-depth understanding of the effectiveness of these methods, we conducted large-scale experiments with as many as \responseline{twelve LLMs on both NLP (\ie, question answering, text summarization, machine translation) and software programming (\ie, code generation) tasks to analyze the correlation between uncertainty estimation results and LLMs performance. The models comprise three from \textit{MetaAI}, four from \textit{OpenAI}, one from \textit{Google}, one from \textit{Microsoft}, one from \textit{Alibaba}, one from \textit{DeepSeek}, and one from \textit{BigCode}. We also evaluated five older LLMs (\eg, GPT-2, LLaMA, Codegen, Incoder, and Santacoder), with the corresponding results available on our website. In total, we have evaluated 17 LLMs} The overall workflow of our work is shown in Fig.~\ref{fig:overall_workflow}. In particular, we investigate the following research questions:


\begin{itemize}[leftmargin=*]

    \item {\bf RQ1:} \rqone
    
    \item {\bf RQ2:} \rqtwo

    \item {\bf RQ3:} \rqthree
    
    \item {\bf RQ4:} \rqfour

\end{itemize}

Our findings validate that uncertainty measurement can, to an extent, be helpful in detecting erroneous responses in general NLP tasks. Additionally, it has also shown to be promising as an indicator for pinpointing faulty programs produced by LLMs.
{Even though, these methods might fall short in detecting nuanced errors made by high-performance commercial models. They seem better suited for filtering out more blatant mistakes.}
Based on the results, we further discuss the insights from our study and highlight a few potential research directions of leveraging uncertainty estimation to enhance the trustworthiness of LLMs for real-world applications across domains. 
\emph{First}, research efforts are needed with novel uncertainty estimation techniques exclusively for LLMs to better fit the corresponding diverse task-handling ability.
{\emph{Second}, we observe that different LLMs can sometimes display markedly distinct uncertain behaviors. Consequently, even though these methods are inherently model-agnostic, stakeholders may need to undertake model-specific optimizations to achieve enhanced performance. \emph{Furthermore}, we observe that the prompt template used in the reinforcement learning from human feedback (RLHF~\cite{ouyang2022training}) could potentially impact the accuracy of uncertainty estimation.}

The contributions of this paper are summarized as follows:

\begin{itemize}[leftmargin=*]
    
    \item We collected and implemented \emph{twelve} different uncertainty estimation methods that are successfully adapted to enable the analysis of LLM, which are also applicable to {both open-source and closed-source LLM models across different downstream tasks in the grey-box setting.}
    
    \item We conducted a large-scale evaluation with nine LLMs on \emph{six} tasks from \emph{four} different domains.
    
    \item We provided an in-depth analysis of the challenges in existing uncertainty methods for LLMs and distilled a set of implications and future opportunities toward reliable and trustworthy LLMs.
    \item Our toolkit, encompassing the dataset, LLM inference, and uncertainty measurement protocols, will be made available for future research endeavors.

\end{itemize}

\noindent \textbf{The Contributions to the Software Engineering Field.} LLMs have revolutionized various aspects of software engineering~\cite{hou2023large, fan2023large}, including but not limited to automated code generation~\cite{fried2023incoder, roziere2023code, allal2023santacoder, nijkamp2022codegen}, software testing~\cite{gu2023llm, kang2023large, lemieux2023codamosa, liu2023fill}, debugging~\cite{tian2024debugbench}, program repair~\cite{first2023baldur, wei2023copiloting}, and document generation~~\cite{geng2024large}. While LLMs can serve as a critical core for many new-era AI-enabled intelligent systems in the software engineering domain, their black-box nature and inherent uncertainties pose challenges for them to be applied in the real world in a transparent, reliable, safe, and secure way. It is thus urgent to investigate and explore effective quality assurance methods. 
Measuring uncertainty and taking it as an indicator of AI models' reliability has been studied extensively in the SE community~\cite{zhang2020towards, OOD2, weiss2023uncertainty}. While promising, most of them focus on classification tasks with relatively simple neural architectures. On the contrary, we initialize a very early stage study on autoregressive, large-scale language models and perform various uncertainty measurements across a wide spectrum of tasks. 

We further provide more supplementary results and details as well as the source code to reproduce our study at our website: \emph{\href{https://sites.google.com/view/llm-uncertainty}{https://sites.google.com/view/llm-uncertainty}}.




\section{Background and Related Work}
\label{sec:background}

\begin{figure*}[h   ]
    \centering
    \includegraphics[width=0.99\linewidth]{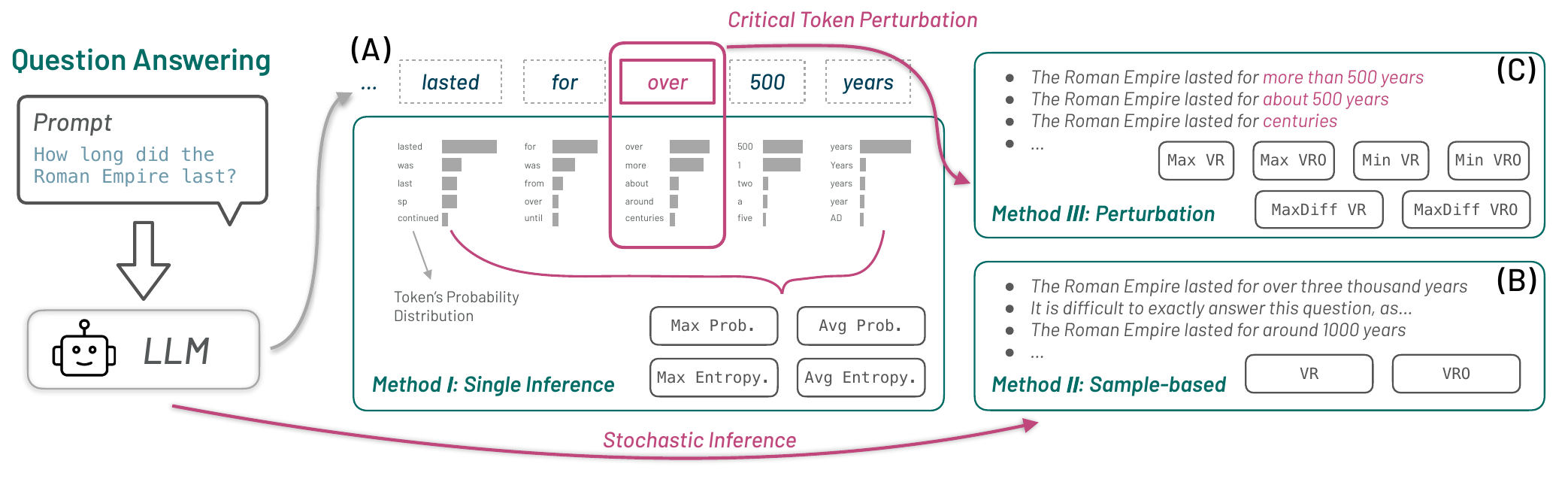}
    \caption{A running example of how different uncertainty estimation methods work for a QA problem with GPT3~\cite{brown2020language}.}
    \label{fig:uncertainty_calculation}
\end{figure*}

\subsection{Large Language Models}
\label{subsec:llms}

In general, a \textit{language model} models sequences of words as a probability distribution, which can be further used to generate coherent and contextually relevant text via conditioning based on a given prompt. Representative traditional language models include HMM (hidden Markov model)~\cite{kuhn1994ergodic}, $n$-gram~\cite{brants2007large}, and RNN (recurrent neural networks)~\cite{mikolov2010recurrent}. Recently, a specific type of neural network architecture, \ie, Transformer~\cite{vaswani2017attention}, has achieved attractive performance on language modelling. Large language models (LLMs) now typically refer to those Transformer-based language models pre-trained with large-scale text corpus and billions of parameters. LLMs have also achieved promising performance in many downstream tasks, \eg, text classification~\cite{howard2018universal}, text summarization~\cite{liu2019text}, and machine translation~\cite{yang2020towards}.

Based on different Transformer architectures and pre-training tasks, LLMs largely fall into three categories: \textit{encoder-only}, \textit{encoder-decoder}, and \textit{decoder-only}. \textit{Encoder-only} LLMs also refer to masked language models (MLM), which are pre-trained through masking a certain portion of tokens (\eg, 15\%) in a sequence. The training objective is to predict those masked tokens correctly. Representative \textit{encoder-only} LLMs include BERT~\cite{kenton2019bert}, RoBERTa~\cite{liu2019roberta}, GraphCodeBERT~\cite{guo2020graphcodebert}, etc. Different from \textit{encode-only} LLMs, \textit{encoder-decoder} LLMs, such as BART~\cite{lewis2020bart}, CodeT5~\cite{wang2021codet5}, are pre-trained through masked span prediction (MSP). \textit{Encoder-decoder} LLMs first learn a representation from input prompts before decoding into another sequence. They are thus usually trained for sequence-to-sequence purposes. Lately, \textit{decoder-only} LLMs have become the mainstream of LLMs research due to their training efficiency and scalability for large-scale datasets and complex model architectures (\ie, billions of model parameters). \textit{Decoder-only} LLMs are autoregressive models. Their training objective is to predict the next token given all previous (left-only) tokens. GPT-based (generative pre-trained transformers) models (\eg, GPT2~\cite{radford2019language}, GPT3~\cite{brown2020language}, LLaMA~\cite{touvron2023llama} and LlaMA-2~\cite{touvron2023llama2}) all belong to this category. In this work, we mainly focus on \textit{decoder-only} LLMs since they have SOTA performance. We further detail subject LLMs in our study in Sec.~\ref{subsec:subject_llms}.

Though LLMs are pre-trained without specific tasks in mind, they can often be used for downstream tasks in two ways, \ie, through (1) prompting and (2) fine-tuning. Prompting refers to the process of in-context learning that ``teaches'' an LLM to solve a specific task by injecting certain knowledge and instructions into the input prompts. Different from prompting, fine-tuning is the process of updating an LLM's neural network weights through a supervised learning strategy for certain tasks (\eg, text classification). Recently, \textit{decoder-only} LLMs have also been used with reinforcement learning from human feedback (RLHF) to improve their performance in understanding complex input prompts and following human instructions. As a result, ChatGPT (GPT3.5 with RLHF)~\cite{chatgpt2023} has shown superior performance in solving various complex tasks (\eg, program synthesizes~\cite{dong2023self}, program repair~\cite{xia2023keep}) by only following the user's instructions and feedback through a dialogue. In our study, we mainly consider LLMs through prompting (that is widely used in practice) and focus on how to estimate an LLM's uncertainty in a ``black-box'' way (\ie, only accessing the model's prediction output probabilities). We introduce our experiment settings in Sec.~\ref{subsec:tasks}.

\subsection{Risk Assessment for ML Models}
\label{subsec:risk_assessment}

Machine learning (ML) models, especially deep learning ones, are known to be notoriously hard to interpret due to their complexity and opacity. Using ML models without appropriate risk assessment could potentially lead to particular concerns and threats regarding trustworthiness, \eg, safety~\cite{varshney2017safety}, security~\cite{barreno2010security}, and ethics~\cite{lo2020ethical}. So far, some risk assessment techniques for general machine learning models have been proposed to reduce the impacts of these concerns~\cite{yang2021generalized, rabanser2019failing, rahmati2019predicting, ning2019optimization, ashukha2020pitfalls}. Among them, there are two representative categories of risk assessment techniques: (1) data distribution analysis and (2) uncertainty estimation. 

The data-driven nature of ML models requires developers to take data into account, especially when test data could be much more different compared with the training data in terms of their distribution. Data distribution analysis, including detecting distribution shift~\cite{rabanser2019failing} and out-of-distribution samples~\cite{hsu2020generalized}, is proposed to identify such differences and avoid potential risks on unseen data. However, data distribution analysis usually requires access to training data, which is often not feasible for LLMs trained on either huge data corpus or private data corpus. Therefore, in this work, we propose to focus on \textit{uncertainty estimation}.

In general, uncertainty estimation aims to measure an ML model's confidence level of a certain prediction. There are two main types of uncertainty in an ML model's predictions: \textit{aleatoric} uncertainty and \textit{epistemic} uncertainty~\cite{kendall2017uncertainties}. \textit{Aleatoric} uncertainty refers to the uncertainty that arises from observations (\eg, sensor noises in an ML model for autonomous driving). By contrast, \textit{epistemic} uncertainty accounts for uncertainty in an ML model's parameters. Insufficient knowledge of an ML model (\eg, lack of a specific type of training data) usually leads to high \textit{epistemic} uncertainty. In this paper, we mainly discuss estimating \textit{epistemic} uncertainty for LLMs. Uncertainty estimation roughly falls into four categories~\cite{gawlikowski2021survey}: (1) \textit{single deterministic methods}~\cite{oberdiek2018classification}, (2) \textit{ensemble methods}~\cite{lakshminarayanan2017simple}, (3) \textit{Bayesian methods}~\cite{barber1998ensemble}, and (4) \textit{test-time augmentation methods}~\cite{lyzhov2020greedy}. \textit{Single deterministic methods} calculate prediction uncertainty based on one forward pass within a deterministic ML model. \textit{Ensemble methods} estimate uncertainty based on a set of different ML models' output. By contrast, \textit{Bayesian methods} only leverage the inherent stochasticity of an ML model (\eg, dropout layer in deep neural networks~\cite{gal2016dropout}). \textit{Test-time augmentation methods} are model-agnostic, which augment the input data at test-time to measure a model's prediction uncertainty~\cite{wang2019aleatoric}. Since we focus on the risk assessment for one standalone LLM, \textit{ensemble methods} are excluded from our study. We detail the uncertainty estimation methods used in this work in Sec.~\ref{sec:uncertainty}.

In addition to the aforementioned general risk assessment techniques, there are also a few works specified for risk assessment of LLMs~\cite{huang2023survey, chen2023robust, jang2023consistency,xiao2021hallucination,malininuncertainty,kuhn2023semantic,manakul2023selfcheckgpt}. The most related works are those proposed for uncertainty estimation of LLMs~\cite{xiao2021hallucination,malininuncertainty,kuhn2023semantic,manakul2023selfcheckgpt}. Xiao \etal leverage \textit{ensemble methods} to measure the natural language generation model's uncertainty and detect potential hallucinations~\cite{xiao2021hallucination}. Similarly, Malinin \etal propose a unified uncertainty estimation method for autoregressive structured prediction tasks based on \textit{ensemble methods}~\cite{malininuncertainty}. To overcome the challenge of capturing ``semantic equivalence'' in natural language, Kuhn \etal propose \textit{semantic entropy} that incorporates linguistic invariances created by shared meanings~\cite{kuhn2023semantic}. Recently, Manakul \etal propose SelfCheckGPT, a black-Box hallucination detection method based on token-level prediction likelihood and entropy~\cite{manakul2023selfcheckgpt}. In light of the limitations of these works, our work is the first work that is not limited to specific natural language or tasks by covering twelve uncertainty estimation methods. Furthermore, our study investigates the role of uncertainty estimation with extensive experiments with nine LLMs and six tasks, providing insights and evidence for its effectiveness as the risk assessment technique for LLMs.

\section{Uncertainty Estimation for LLMs}
\label{sec:uncertainty}

In this section, we first discuss the problem scenario in our study, including the corresponding assumptions. Then, we introduce our twelve uncertainty estimation techniques (three categories) based on the number of inferences required.



\subsection{Problem Scenario}
Given an input prompt $X=[x_1,x_2,\dots,x_n]$ ($x_i$ denotes $i$th input token), an LLM $f$ with pre-trained weights $w$ generates another sequence $Y=[y_1, y_2, \dots, y_m]$ ($y_j$ denotes $j$th generated token) through a decoding process: $y_j = f\left([X, y_1, y_2, \dots, y_{j-1}]|w\right)$


An uncertainty estimation method $g$ is to calculate a score $u$ regarding the uncertainty of $Y$.


Though an LLM can be regarded as an ML model, it is limited by the inherent properties of some existing uncertainty estimation methods. Following we discuss the unique characteristics of LLMs and challenges in uncertainty estimation for LLMs compared with other ML models.


\begin{itemize}[leftmargin=*]
  \item {\bf Complexity.} The state-of-the-art (SOTA) LLMs are usually pre-trained with billions of parameters (\eg, GPT-3~\cite{brown2020language} model has 96 layers with 6.7 billion parameters). Therefore, ``white-box'' analysis for interpreting LLMs (\eg, inspecting neuron activation~\cite{alammar-2021-ecco}, inspecting attention values~\cite{galassi2020attention, vig2019analyzing}) that requires both significant manual and computational efforts is not feasible. 
  
  \item {\bf Opacity.} There is also a lack of opacity in SOTA LLMs. First, the SOTA LLMs are usually trained with large-scale text corpus, where such data can be either publicly available or from private sources. Therefore, risk assessment techniques that require access to training data (\eg, OOD detection) can not be used in our context. Additionally, some of the SOTA LLMs are potentially proprietary assets for a company (\eg, GPT-3~\cite{brown2020language}), where one can only access the inference results through provided APIs.

  
  \item {\bf Task diversity.} Though the usage of LLMs can be described in a general decoding form 
  , tasks that LLMs can solve are of greater diversity. 
  Notably, LLMs can be used for user-defined tasks through prompting/few-shot learning (Sec.~\ref{subsec:llms}). 
  Therefore, uncertainty estimation methods that are proposed for a specific narrow domain (\eg, text classification) are hard to be used as a general risk assessment technique for LLMs. 
\end{itemize}

 \begin{figure}[tbp]
      \centering
      \includegraphics[width=0.95\linewidth]{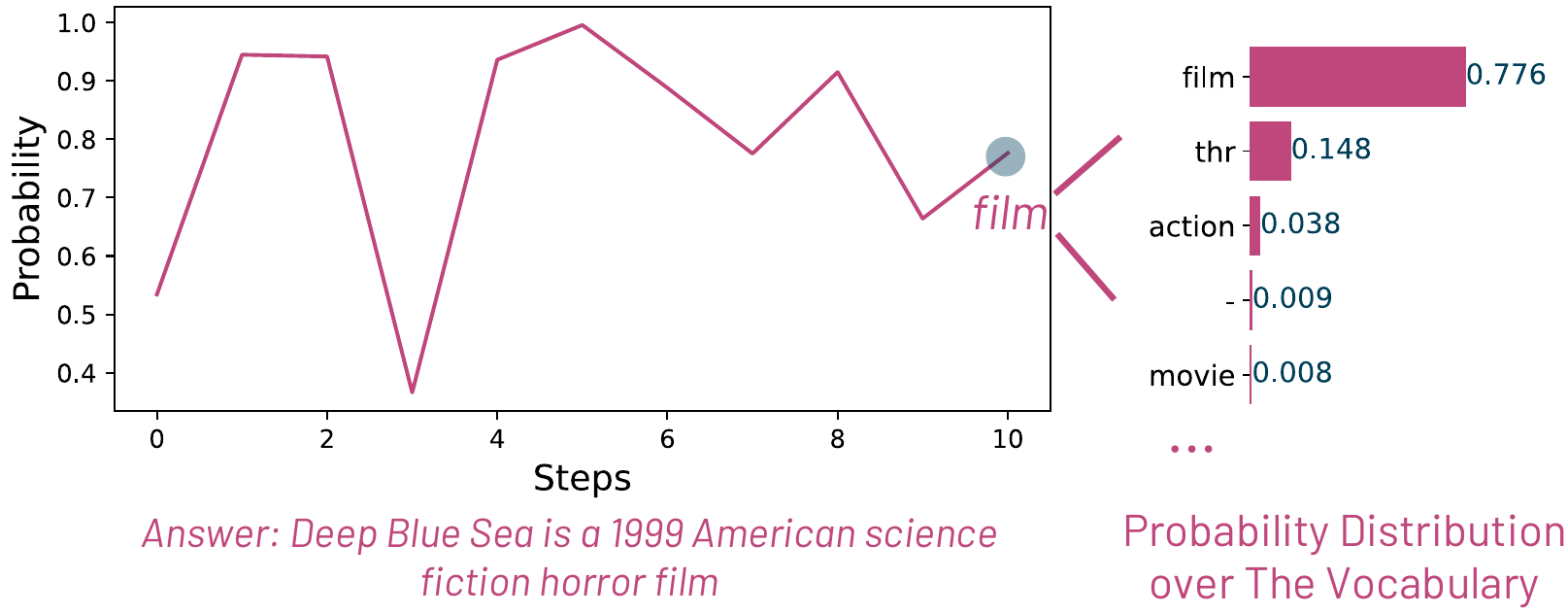}
          \caption{The available information we can get from the LLM in our setting. This is an illustration of how \textit{LLAMA-7B} answers the question \textit{``when did the movie deep blue sea come out?''}.}
      \label{fig:bg:illustration}
\end{figure}

With these characteristics and challenges in mind, we show the information that we can get from the majority of both open-source and closed-source LLMs in Fig.~\ref{fig:uncertainty_calculation} (A) and explore possible existing solutions in this grey box setting for a general usage risk assessment. We summarize the necessary information needed in this study in Fig.~\ref{fig:bg:illustration}. 
In most cases, we are only able to obtain the output tokens as well as the probability distribution over the vocabulary for each token. 
Specifically, for closed-source models such as GPT-3~\cite{brown2020language} and GPT-3.5~\cite{gpt-3.5}, one is only able to obtain the top-$k$ probabilities (\ie, $k$ possible tokens with the highest probabilities) for each predicted token. Formally, we can formulate the uncertainty estimation in our study as
\begin{equation}
    \small
    u = g(f(\cdot), X, Y, P),
\end{equation}
where $P$ is either the probability distribution or top-$k$ probabilities for each token.

Finally, we select twelve uncertainty estimation methods covering \textit{single deterministic methods}, \textit{Bayesian methods}, and \textit{test-time augmentation methods} (see discussion in Sec.~\ref{subsec:risk_assessment}). We categorize our uncertainty estimation methods based on the number of inferences required and detail them in the following.



\subsection{Single-inference Uncertainty Estimation}

Single-inference uncertainty estimation methods can be seen as \textit{single deterministic methods} in Sec.~\ref{subsec:risk_assessment}. These methods usually calculate an ML model's confidence based on the probability distribution of the prediction~\cite{uncertainty_naive1, uncertainty_naive2}. In particular, such methods are usually used for classification tasks. Though LLMs can be used in various different tasks, the generation of each token 
can still refer to a classification problem (\ie, choose one token from the entire vocabulary). To aggregate the uncertainty information obtained at the token level, Manakul \etal \cite{manakul2023selfcheckgpt} propose four different metrics to aggregate token-level uncertainty into sentence level. 
%
%
%

In particular, a sentence-level uncertainty score can be obtained by taking either the maximum or average of the likelihood $-\log p$ in a sentence:
\begin{equation}
    \label{eq:max_prob}
    \small
    Max(-\log p)_i = \underset{j}{max} ( -\log p_{ij}),
\end{equation}
\begin{equation}
    \label{eq:avg_prob}
    \small
    Avg(-\log p)_i = -\frac{1}{J}\sum_{j} \log p_{ij},
\end{equation}
where $p_{ij}$ is the probability of a token at position $j$ of a sentence $i$.
Additionally, one can also replace the likelihood $-\log p$ with the entropy $\mathcal{H}$:
\begin{equation}
    \label{eq:max_entropy}
    \small
    Max(\mathcal{H})_i = \underset{j}{max} [\mathcal{H}_{ij}],
\end{equation}
\begin{equation}
    \label{eq:avg_entropy}
    \small
    Avg(\mathcal{H})_i = \frac{1}{J}\sum_{j} \mathcal{H}_{ij},
\end{equation}
where $\mathcal{H}_{ij}$ is the entropy of this token's probability distribution over the vocabulary.


After obtaining sentence-level uncertainty estimation, one can further calculate the passage-level uncertainty score by taking the average over all sentence-level uncertainty scores.
%
In this study, we use the metrics discussed above as the {\bf single-inference uncertainty estimation methods} and denote them as \textbf{Max Prob}, \textbf{Average Prob}, \textbf{Max Ent} and \textbf{Average Ent}.


\subsection{Multi-inference Uncertainty Estimation}

Multi-inference uncertainty estimation methods leverage the stochastic in either a model's parameters (\eg, \textit{Bayesian methods}) or data (\eg, \textit{test-time data augmentation methods}) to collect a set of non-deterministic predictions. A model's prediction uncertainty is then estimated as the divergence among those predictions. 

\subsubsection{Metrics} Two metrics were widely used to measure such divergence: (1) variation ratio (VR) and (2) variation ratio for original prediction (VRO)~\cite{gal2016dropout, Gal2016UncertaintyID, zhang2020towards}. Originally, both metrics are defined for a classification problem. Wang~\etal~\cite{wang2022exploratory} extend the definitions of \textit{VR}/\textit{VRO} and show that they are still effective in tasks other than classification. We introduce these two metrics in the following:
%
%
\begin{equation}
    \small
    \label{eq:evr}
    VR = 1 - \frac{\sum_{i=1}^T w*\frac{\sum_{j=1, j \neq i}^{j=T} (1 - dist(p_i, p_j))}{T-1}}{T},
\end{equation}
\begin{equation}
    \small
    \label{eq:evro}
    VRO = 1 - \frac{\sum_{i=1}^T (1 - dist(p_i, p_{L_M}))}{T},
\end{equation}
where $T$ is the number of inferences, $dist(\cdot)$ denotes the distance function between two outputs. $p_i$ and $p_j$  are the inference result at $i$th and $j$th inference. $p_{L_M}$ is the prediction result from the original model $M$. $w$ denotes a weight matrix.

\begin{figure}[h!]
    \centering
    \includegraphics[width=0.99\linewidth]{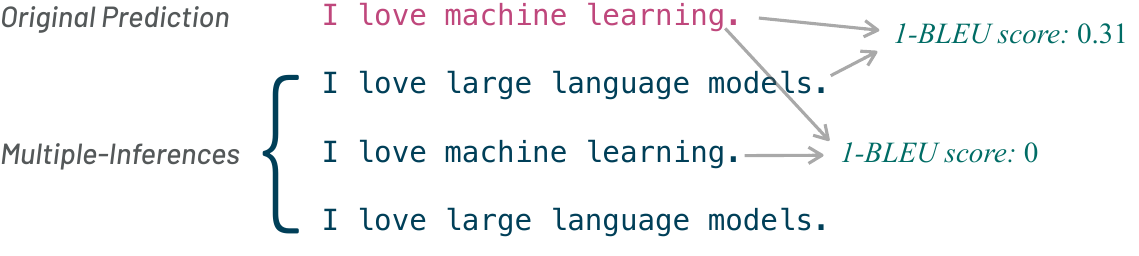}
    \vspace{-5pt}
    \caption{An example of multiple inferences with LLMs.}
    \label{fig:vr_vro}
\end{figure}

We now use an example to demonstrate the calculation of \textit{VR}/\textit{VRO} in Fig.~\ref{fig:vr_vro}. Suppose we have three different generations given stochasticity in either model or data. $(1-BLEU)$ score~\cite{papineni2002bleu} is used as $dist(\cdot)$ to measure the differences between two sentences~\footnote{A higher $(1-BLEU)$ indicates a lower similarity between two texts.}. Suppose we use identical weights for equation~\ref{eq:evr}, then:
\begin{equation}
    \small
    VR= 1 - \frac{\frac{0.69+1}{2} + \frac{0.69+0.69}{2} + \frac{0.69+1}{2}}{3} \approx 0.21.
\end{equation}
Similarly, VRO is
\begin{equation}
    \small
    VRO = 1 - \frac{0.69+0.69+1}{3}\approx0.21.
\end{equation}

\subsubsection{Stochastic inference} To enable the stochasticity of an ML model, one popular way is to execute it with the dropout layer(s) enabled at test-time~\cite{gal2016dropout}. However, this is not feasible given that we focus on ``black-box'' analysis. We introduce two different ways to enable stochasticity in our study: (1) {\bf sample-based method} (\textit{Bayesian methods}) and (2) {\bf perturbation-based method} (\textit{test-time augmentation methods}).

\vspace{1mm}
{\noindent \bf Sample-based method.}
Randomness exists in LLMs' generation process. Specifically, by controlling the parameter \textit{temperature} ($t$), an LLM's generation can be either deterministic or non-deterministic. When $t$ is 0, an LLM will always choose the token with the largest probability (\ie, greedy decoding). Thus the generation will be deterministic. When $t>0$, an LLM will randomly choose a token as long as its probability is larger than a threshold. A higher $t$ will lead to a larger randomness in an LLM's generation. In this study, we set $t>0$ to enable an LLM's stochasticity and generate non-deterministic outputs. We refer to this method as \textit{sample-based method}. We denote the usages of two different metrics as \textbf{Sample VR} and \textbf{Sample VRO}. We detail the choices of $t$ in Sec.~\ref{subsec:experiment_settings}.


\vspace{1mm}
{\noindent \bf Perturbation-based method.} An LLM's stochasticity could also be enabled when changing the input prompt or perturbing a generated token. This also refers to the \textit{test-time augmentation methods}. Intuitively, along the chain of the token generation, any perturbation of a generated token can affect the proceeding tokens' generation 
, and possibly result in two semantically different texts. We refer to an example in Fig.~\ref{fig:uncertainty_calculation}~(C), where LLM answers the question, \textit{``How long did the Roman Empire last?''} Changing the word \textcolor{amii_magenta}{\textit{over}} in the generation results could yield a completely different result, \textit{``lasted for \underline{centuries}.''} This phenomenon shows an intriguing property of LLMs: their stochastic nature along the prediction chain with respect to perturbation. 

Therefore, in this study, we propose an adaptive method to perturb ``critical'' token(s) during an LLM's generation process. Prior work has shown that tokens with high entropy in a sentence usually convey more information from the information theory perspective~\cite{meister2023locally}. Consequently, perturbations at such points could possibly lead to a more pronounced change compared with others. 
To explore the effectiveness of using such perturbation for measuring uncertainty, we define three types of interest points to perturb: (1) the point with the highest entropy; (2) the point with the lowest entropy; (3) the point that gains the maximum entropy from the previous token in a response. In our study, we select the token that matches one of these three types of points and replace it with $k$ other tokens with the highest probabilities. 
We name this method \textit{perturbation-based method} and denote six different variants as: \textbf{Max VR}, \textbf{Max VRO}, \textbf{Min VR}, \textbf{Min VRO}, \textbf{MaxDiff VR}, and \textbf{MaxDiff VRO}. We justify the choice of $k$ in Sec.~\ref{subsec:experiment_settings}.


\section{Study Design}
\label{sec:study_design}

\begin{table*}[t]
    \setlength{\tabcolsep}{5pt}
    \responseref{}
    \centering
    \scriptsize
    
    \renewcommand{\arraystretch}{1.1}
    
    \caption{Subject open-source LLMs in our study.}
    \begin{tabular}{lcccccccc}
        \toprule
         {\bf LLMs} & {LLaMA2~\cite{touvron2023llama2}} & {LLaMA3~\cite{dubey2024llama}} & {Gemma2~\cite{team2024gemma}} & {Phi3~\cite{abdin2024phi}}  & {CodeQwen1.5~\cite{qwen}} & {DeepSeek Coder~\cite{guo2024deepseek}} & {Starcoder 2~\cite{lozhkov2024starcoder}} & {Code Llama~\cite{roziere2023code}} \\
         \midrule
         {\bf Model Size} & 6.7B & 8B & 9B & 7B   & 7B & 6.7B & 7B & 6.7B\\
         {\bf Training Data} & 2T tokens & 15T tokens & 8T tokens& 4.8T tokens  & 2.4T + 90B & 2T & 3.5T & 2T + 620B\\
         
         {\bf Domain} & Text & Text & Text & Text &  Code & Code & Code & Code \\
         
         {\bf Provider} & MetaAI & MetaAI & Google & Microsoft  & Alibaba & DeepSeek & BigCode & MetaAI\\
         
         \bottomrule
    \end{tabular}
    \label{tab:subject_llms_1}
\end{table*}

\begin{table}[t]
    \setlength{\tabcolsep}{3pt}
    \responseref{}
    \centering
    
    \scriptsize
    
    \renewcommand{\arraystretch}{1.1}
    
    \caption{Subject closed-source LLMs in our study.}
    \begin{tabular}{lccccc}
        \toprule
         {\bf LLMs} & {GPT-3~\cite{brown2020language}} &{GPT-3.5~\cite{gpt-3.5}} & {GPT4o~\cite{gpt-4o}} & {GPT4o mini~\cite{gpt-4o-mini}}\\
         \midrule
         {\bf Model Size} & 6.7B & 175B & Unknown & Unknown\\

         {\bf Training Data} & 570 GB & Unknown & Unknown& Unknown\\

         {\bf Domain} & Text & General & {\tiny Multimodal General} & {\tiny Multimodal General} \\

         {\bf Provider} & OpenAI & OpenAI & OpenAI & OpenAI\\

         \bottomrule
    \end{tabular}
    \label{tab:subject_llms_2}
\end{table}

In this section, we introduce our study design and methodology to answer our research questions in Sec.~\ref{sec:introduction}.

\subsection{Subject LLMs}
\label{subsec:subject_llms}

In order to provide a thorough evaluation of the effectiveness of uncertainty measurement as the risk assessment for LLMs in both natural language and domain-specific (\ie, code-generation) problems, we have chosen a broad spectrum of representative models. We selected these models considering their availability, diversity, and computational requirements. \responseline{Four open-source LLMs for NLP tasks, four closed-source general LLMs, and four LLMs specified for code generation are selected. We present our selected LLMs in Table~\ref{tab:subject_llms_1} and Table~\ref{tab:subject_llms_2}. We additionally performed experiments on older LLMs including GPT-2-XL~\cite{radford2019language}, LLaMA-7B~\cite{touvron2023llama}, Incoder~\cite{fried2023incoder}, SantaCoder~\cite{allal2023santacoder} and CodeGen~\cite{nijkamp2022codegen}. Due to space limits, we left related results on our website.}

\vspace{-2mm}

\subsection{Tasks}
\label{subsec:tasks}

\begin{table}[!tb]
\caption{The collected NLP and programming tasks.}
\centering
\setlength{\tabcolsep}{4pt}
\renewcommand{\arraystretch}{1.1}
\scriptsize
\renewcommand{\arraystretch}{1.1}
\label{tab:tasks}
    \begin{tabular}{llcc}
    \toprule 
    \textbf{Dataset} & \textbf{Task Domain}  & \multicolumn{1}{p{1.4cm}}{\centering \textbf{Size} \\ }  & \multicolumn{1}{p{2cm}}{\centering \textbf{$dist(\cdot)$ metrics} \\ \textbf{in \textit{VR}/\textit{VRO}}}  \ \\ \midrule
    {\bf Eli5-Category} & Question Answering  & 5,411 & F1 score~\cite{yang-etal-2015-wikiqa} \\
    {\bf Wiki-QA} & Question Answering  & 243 & F1 score~\cite{yang-etal-2015-wikiqa} \\
    {\bf CNN/Daily Mail} & Text Summarization  & 11,490 & ROUGE-L~\cite{lin-2004-rouge} \\
    {\bf WMT 2014}    & Machine Translation & 3,003 & BLEU~\cite{papineni2002bleu}\\
    {\bf MBPP} & Code Generation & 500 & CodeBLEU~\cite{ren2020codebleu}\\
    {\bf HumanEval} & Code Generation & 164 & CodeBLEU~\cite{ren2020codebleu}\\
    \bottomrule
    \end{tabular}
    
\end{table}

To comprehensively understand the effectiveness of selected uncertainty estimation methods, we selected a set of diverse and challenging tasks as a benchmark (Table.~\ref{tab:tasks}). Specifically, our evaluation covers four NLP tasks from three different domains (i.e., question answering, text summarization, and machine translation). Furthermore, our evaluation also includes two different code-generation tasks. Below, we introduce the tasks of different domains.

{\noindent \textbf{Question Answering (QA)}} 
requires an LLM with the capability of understanding users' intentions, extracting learned knowledge, and organizing responses. In this study, we selected two different benchmarks for QA: \textit{Eli5-category}~\cite{eli5-category} and \textit{Wiki-QA}~\cite{yang-etal-2015-wikiqa}. \textit{Eli5-category} is a collection of 5,411 questions and answers gathered by and {obtained from pushshift.io}. 
\textit{Wiki-QA} is a collection of questions extracted from \textit{Bing} query logs with Wikipedia summaries as answers. We selected 243 instances after the de-duplication and removal of questions without answers. 

\vspace{1mm}
{\noindent \textbf{Text Summarization}} aims to condense a long text into a concise short summary. Different from QA, 
text summarization focuses on benchmarking an LLM's capabilities of extracting critical information from a long paragraph of text. We use the CNN/Daily Mail dataset~\cite{nallapati2016abstractive} as the benchmark for text summarization. It comprises 11,490 news articles from CNN and the Daily Mail, with summaries obtained through the concatenation of highlighted sentences as composed by the original authors. Note that during the evaluation, we add a prompt {\ttfamily TL;DR} after the input text to enable an LLM's in-context learning ability for summarization~\cite{radford2019language}.

\vspace{1mm}
{\noindent \textbf{Machine Translation}} is another fundamental task to benchmark language models. In this study, we use the WMT 2014 dataset~\cite{bojar-etal-2014-findings}, including 3,003 pairs of French-English transcripts, to evaluate LLMs. Similar to text summarization, {We used the prompt template from the GPT-3 paper~\cite{brown2020language}, where a random example translation is presented in the input for in-context learning.} 



\vspace{1mm}
{\noindent \textbf{Code Generation}} requires an LLM to understand both natural language (e.g., task description) and programming languages (e.g., formal structures and precise syntax). \responseline{We chose it as a representative task for evaluating LLMs’ coding abilities because (1) it is a comprehensive task that challenges the models' advanced understanding and reasoning capabilities, (2) it has been selected as one of the major tasks to benchmark LLMs' code abilities~\cite{roziere2023code, qwen, guo2024deepseek, lozhkov2024starcoder}, and (3) it can be automated and objectively evaluated through test cases.} We select two datasets as benchmarks for code generation: HumanEval~\cite{chen2021evaluating} and MBPP~\cite{austin2021program}.
HumanEval consists of 164 programming problems with manually written test cases released by OpenAI. MBPP includes 1,000 crowd-sourced Python programming problems with entry-level difficulties. 



\subsection{Evaluation Metrics}

\vspace{1mm}
{\noindent \bf NLP Tasks.} We use \textit{semantic distance} to measure the performance of LLMs' generation for NLP tasks. Specifically, we first embed the text using sentence-transformer~\cite{reimers-2019-sentence-bert} and compute the cosine distance of the embeddings. A higher cosine distance value indicates a greater level of similarity. This metric can be generalized to different NLP tasks regardless of the length and form of generation. Note that while there are some other metrics specialized for each NLP task, which relies on string-matching (\eg, F1 score for text summarization), we argue two major shortcomings exist when using such metrics. First, string-matching might not accurately capture the divergence between the model's output and the ground truth (\eg, when the output and ground truth are semantically equivalent while lexically different). Additionally, LLMs without fine-tuning might generate responses in a free-form manner that are often longer than the ground truth in, \eg, text summarization and question answering. In this case, even if an LLM answers correctly, metrics based on string matching could still mis-indicate its performance due to the length differences.


\vspace{1mm}
{\bf \noindent Code Generation Task.} {Different from NLP tasks, it is relatively easy to assess the quality of generated code. We introduce a quality score $Q$ given by $Q = (Q_{syntax} + Q_{semantics}) / {2}$. $Q_{syntax} = 1$ if the generated code is syntactically correct (\ie, it can be executed by the interpreter). Meanwhile, $Q_{semantics}$ is the proportion of test cases the generated code passes.}



\subsection{Experiment Settings}
\label{subsec:experiment_settings}

{\noindent \bf Uncertainty measurement.} We investigate twelve uncertainty estimation methods in our experiments: four \textit{single-inference} methods, two \textit{sample-based} methods, and six \textit{perturbation-based} methods. We set the number of inferences $T$ to $5$ for both \textit{sample-based} and \textit{perturbation-based} methods. This is for a fair comparison between these two methods since closed-source LLMs (\ie, GPT-3, GPT-3.5) only provide access to top-5 tokens for each token's generation. Therefore, only 5 inferences can be obtained with the \textit{perturbation-based} method on closed-source LLMs. For temperature $t$, we follow the previous work and set $t$ to $0.7$ when enabling on an LLMs' stochasticity~\cite{chen2023teaching, cobbe2021training}. 

For distance metric in \textit{VR} and \textit{VRO} (\ie, $dist(\cdot)$ in Eq.~\ref{eq:evr}\&\ref{eq:evro}), we consider both (1) general metrics based on embeddings' cosine distance and (2) task-specific metrics. For general metrics, we use \textit{all-mpnet-base-v2}~\cite{mpnet-base} for natural language tasks and \textit{codebert-base}~\cite{guo2020graphcodebert} for code data. For task-specific metrics, we consider the following choices in Table~\ref{tab:tasks}. 

\begin{table*}[htbp]
  \centering
  \caption{Pearson correlation coefficients between uncertainty scores and LLMs' performance on four NLP tasks. The results of VR/VRO are presented as ``{\em cosine distance-based (task-specific distance-based)}.'' Highest correlations from different categories are ranked and highlighted as \colorbox{top1Color}{\textit{top-1}}, \colorbox{top2Color}{\textit{top-2}}, and \colorbox{top3Color}{\textit{top-3}}. 
  }
  \renewcommand{\arraystretch}{1.1}
  \begin{subtable}[h]{\textwidth}
    \setlength{\tabcolsep}{10pt}
    \centering
    \responseref{}
    \scriptsize
        \begin{tabular}{llcccccll}
        \toprule
        \multirow{2}{*}{\textbf{Dataset}}&\multirow{2}{*}{\textbf{LLM}} & \multicolumn{4}{c}{{\bf Single-inference Method}} && \multicolumn{2}{c}{{\bf Sample-based Method}}\\
        \cmidrule{3-6} \cmidrule{8-9}
        && \textbf{Max Prob} & \textbf{Average Prob} & \textbf{Max Ent} & \textbf{Average Ent} && \textbf{Sample VR} & \textbf{Sample VRO}\\
        \cmidrule{1-6} \cmidrule{8-9}

        \multirow{8}{*}{{CNN/Daily Mail}} & LLaMA2 & -0.096 & -0.109 &-0.139 & -0.110  && -0.370(-0.260) & \toponehl{-0.534}(-0.294)  \\
        
        & LLaMA3 & 0.109 & -0.342 & -0.002 & -0.433                            && \toponehl{-0.643}(\topthreehl{-0.510}) & \toptwohl{-0.590}(-0.307)  \\

        & Gemma2 & -0.331 & -0.336 & -0.211 & -0.317  && -0.272(-0.025) & \toptwohl{-0.486}(-0.204)  \\

        & Phi3 & -0.535 & -0.620 & -0.606 & -0.648  && \toptwohl{-0.709}({-0.651}) & \toponehl{-0.739}(-0.648)  \\
        
        & GPT3 & \toptwohl{-0.231} & -0.202 & -0.145 & -0.170 && \topthreehl{-0.229}(-0.100) & \toponehl{-0.394}(-0.244)  \\
        
        & GPT3.5 & -0.119 & -0.119 & -0.036 & -0.106 && \toponehl{-0.223}(-0.131) &\toptwohl{-0.220}(-0.158)	 \\

        & GPT4o mini& -0.075 & -0.196 & -0.021 & -0.211 && -0.199(-0.187) & -0.177(-0.164) \\

        & GPT4o & \topthreehl{-0.334} & -0.150 & -0.196 & -0.126 && \toponehl{-0.383}(\toptwohl{-0.338}) & -0.324(-0.290) \\

        \cellcolor{lightgray}& \cellcolor{lightgray}LLaMA2 & \cellcolor{lightgray}-0.000 & \cellcolor{lightgray}-0.008&
        \cellcolor{lightgray}0.008 & \cellcolor{lightgray} -0.018 && \cellcolor{lightgray}-0.196(-0.006) & \cellcolor{lightgray}\toponehl{-0.300}(0.024)  \\

        \cellcolor{lightgray}& \cellcolor{lightgray}LLaMA3 & \cellcolor{lightgray}-0.007 & \cellcolor{lightgray}-0.046 & \cellcolor{lightgray}-0.043 & \cellcolor{lightgray}-0.067 && \cellcolor{lightgray}\toptwohl{-0.173}(-0.051) & \cellcolor{lightgray}\toponehl{-0.177}(-0.064)  \\

        \cellcolor{lightgray}& \cellcolor{lightgray}Gemma2 & \cellcolor{lightgray}0.212 & \cellcolor{lightgray}0.195 & \cellcolor{lightgray}0.168 & \cellcolor{lightgray}0.201 && \cellcolor{lightgray}0.001(-0.088) & \cellcolor{lightgray}-0.099(-0.106)  \\

        \cellcolor{lightgray}& \cellcolor{lightgray}Phi3 & \cellcolor{lightgray}-0.007 & \cellcolor{lightgray}-0.008 & \cellcolor{lightgray}-0.044 & \cellcolor{lightgray}-0.027 && \cellcolor{lightgray}-0.002(0.088) & \cellcolor{lightgray}-0.140(-0.013)  \\

        \cellcolor{lightgray}& \cellcolor{lightgray}GPT3 & \cellcolor{lightgray}-0.054 & \cellcolor{lightgray}-0.207 & \cellcolor{lightgray}-0.016 & \cellcolor{lightgray}-0.172 && \cellcolor{lightgray}\topthreehl{-0.236}(-0.084) & \cellcolor{lightgray}\toponehl{-0.382}(-0.140)	  \\
        
         \cellcolor{lightgray}& \cellcolor{lightgray}GPT3.5 & \cellcolor{lightgray}-0.037 & \cellcolor{lightgray}\toptwohl{-0.240} & \cellcolor{lightgray}-0.084 & \cellcolor{lightgray}\toponehl{-0.289} && \cellcolor{lightgray}-0.168(-0.021) & \cellcolor{lightgray}\topthreehl{-0.223}(-0.046)  \\

        \cellcolor{lightgray}& \cellcolor{lightgray}GPT4o mini & \cellcolor{lightgray}-0.200 & \cellcolor{lightgray}\topthreehl{-0.283} & \cellcolor{lightgray}-0.099 & \cellcolor{lightgray}\toptwohl{-0.303} && \cellcolor{lightgray}-0.116(-0.162) & \cellcolor{lightgray}-0.127(-0.064)	  \\

        \multirow{-8}{*}{\cellcolor{lightgray}{Eli5-Category}}& \cellcolor{lightgray}GPT4o & \cellcolor{lightgray}0.019 & \cellcolor{lightgray}\topthreehl{-0.179} & \cellcolor{lightgray}0.007 & \cellcolor{lightgray}\toptwohl{-0.207} && \cellcolor{lightgray}\toponehl{-0.224}(-0.146) & \cellcolor{lightgray}-0.152(-0.118)  \\

        \multirow{8}{*}{{Wiki-QA}} & LLaMA2 & -0.181 & \topthreehl{-0.193} & -0.166 & -0.140 && \toptwohl{-0.281}(-0.206) & \toponehl{-0.374}(-0.184)	\\

        & LLaMA3 & 0.017 & 0.091 & -0.011 & 0.093 && \toptwohl{-0.227}(-0.060) & \toponehl{-0.294}(-0.050)	 \\

        & Gemma2 & 0.226 & 0.265 & 0.259 & 0.300 && -0.031(0.213) & -0.138(0.212)	 \\

        & Phi3 & 0.038 & -0.101 & -0.004 & -0.101 && \toponehl{-0.395}(-0.045) & \toptwohl{-0.370}(0.030)	 \\
        
        & GPT3 & -0.112 & -0.192 & 0.017 & -0.107 && \toptwohl{-0.326}(-0.061) & \toponehl{-0.376}(-0.139)  \\
        
        & GPT3.5 & -0.051 & -0.107 & -0.065 & -0.105 && \toptwohl{-0.401}(-0.085) & \toponehl{-0.425}(-0.079) \\

        & GPT4o mini & 0.032 & 0.031 & -0.018 & 0.011 && \toptwohl{-0.223}(-0.051) & -0.124(0.010) \\

        & GPT4o & 0.080 & -0.052 & 0.006 & -0.083 && \toponehl{-0.222}(-0.074) & \topthreehl{-0.128}(-0.054) \\
      
        \cellcolor{lightgray}& \cellcolor{lightgray}LLaMA2 & \cellcolor{lightgray}0.157 & \cellcolor{lightgray}-0.000 & \cellcolor{lightgray}\toptwohl{0.185} & \cellcolor{lightgray}-0.015 && \cellcolor{lightgray}0.068(0.009) & \cellcolor{lightgray}0.125(\topthreehl{0.163})  \\

        \cellcolor{lightgray}& \cellcolor{lightgray}LLaMA3 & \cellcolor{lightgray}-0.113 & \cellcolor{lightgray}-0.001 & \cellcolor{lightgray}-0.139 & \cellcolor{lightgray}-0.020 && \cellcolor{lightgray}0.006(0.105) & \cellcolor{lightgray}-0.071(-0.001) \\

        \cellcolor{lightgray}& \cellcolor{lightgray}Gemma2 & \cellcolor{lightgray}0.342 & \cellcolor{lightgray}0.153 & \cellcolor{lightgray}0.309 & \cellcolor{lightgray}0.115 && \cellcolor{lightgray}\toptwohl{-0.510}(-0.410) & \cellcolor{lightgray}\toponehl{-0.769}(\topthreehl{-0.485}) \\

        \cellcolor{lightgray}& \cellcolor{lightgray}Phi3 & \cellcolor{lightgray}-0.024 & \cellcolor{lightgray}-0.126 & \cellcolor{lightgray}-0.071 & \cellcolor{lightgray}-0.09 && \cellcolor{lightgray}\toponehl{-0.393}(-0.159) & \cellcolor{lightgray}\toptwohl{-0.392}(\topthreehl{-0.164})\\

        \cellcolor{lightgray}& \cellcolor{lightgray}GPT3 & \cellcolor{lightgray}-0.046 & \cellcolor{lightgray}-0.162 & \cellcolor{lightgray}-0.039 & \cellcolor{lightgray} -0.176 && \cellcolor{lightgray}-0.250(\toptwohl{-0.254}) & \cellcolor{lightgray}-0.251(\toponehl{-0.339})  \\
        
         \cellcolor{lightgray} & \cellcolor{lightgray}GPT3.5 & \cellcolor{lightgray}0.089 & \cellcolor{lightgray}-0.158 & \cellcolor{lightgray}0.050 & \cellcolor{lightgray}{-0.175} && \cellcolor{lightgray}\toptwohl{-0.244}(-0.065) & \cellcolor{lightgray}\topthreehl{-0.233}(-0.017)  \\

        \cellcolor{lightgray} & \cellcolor{lightgray}GPT4o mini & \cellcolor{lightgray} \topthreehl{-0.223} & \cellcolor{lightgray} -0.136 & \cellcolor{lightgray} -0.138 & \cellcolor{lightgray} -0.122 && \cellcolor{lightgray} \toponehl{-0.403}(-0.056) & \cellcolor{lightgray} \toptwohl{-0.394}(-0.095))  \\

        \multirow{-8}{*}{\cellcolor{lightgray}{WMT 2014}} & \cellcolor{lightgray}GPT4o & \cellcolor{lightgray}-0.054 & \cellcolor{lightgray}\topthreehl{-0.273} & \cellcolor{lightgray}-0.06 & \cellcolor{lightgray}-0.265 && \cellcolor{lightgray}\toponehl{-0.558}(-0.209) & \cellcolor{lightgray}\toptwohl{-0.539}(-0.137)  \\
        
        \bottomrule
        \end{tabular}
  \end{subtable}
  \vspace{6pt}
  \vfill
  \begin{subtable}[h]{\textwidth}
  \setlength{\tabcolsep}{9.15pt}
  \centering
  \responseref{}
  \scriptsize
    \begin{tabular}{llcccccc}
        \toprule
        \multirow{2}{*}{\textbf{Dataset}}&\multirow{2}{*}{\textbf{LLM}} & \multicolumn{6}{c}{\textbf{Perturbation-based}}\\
        \cmidrule{3-8}
        && \textbf{Max VR} & \textbf{Max VRO} & \textbf{Min VR} & \textbf{Min VRO} & \textbf{MaxDiff VR} & \textbf{MaxDiff VRO} \\
        \midrule
        
        \multirow{8}{*}{CNN/Daily Mail} & LLaMA2 &  -0.204(-0.161) & \topthreehl{-0.388}(-0.154) & -0.075(-0.055) & \toptwohl{-0.418}(-0.226) & -0.194(-0.152) & -0.378(-0.143)  \\

        & LLaMA3 &  0.027(0.005) & 0.123(0.231) & 0.472(0.385) & 0.181(0.204) & 0.076(0.043) & 0.167(0.256)  \\

        & Gemma2 &  0.405(\toponehl{0.488}) & 0.308(0.435) & 0.296(0.304) & 0.378(0.422) & 0.443(\topthreehl{0.474}) & 0.404(0.461)  \\

        & Phi3 & 0.447(0.555) & 0.579(0.669) & 0.343(0.446) & 0.511(0.598) & 0.437(0.551) & 0.570(\topthreehl{0.659}) \\
        
        & GPT3 &  -0.096(-0.084) & -0.233(-0.209) & -0.001(-0.049) & 0.034(0.041) & -0.119(-0.121) & -0.228(-0.182)	  \\
        	
        & GPT3.5 & 0.064(-0.086) & -0.024(-0.087) & -0.116(-0.176) & -0.068(\topthreehl{-0.177}) & -0.042(-0.138) & 0.072(0.021)  \\

        & GPT4o mini & \topthreehl{-0.227}(\toponehl{-0.288}) & 0.065(0.058) & 0.085(0.020) & 0.058(0.048) & -0.118(\toptwohl{-0.240}) & 0.138(0.111) \\

        & GPT4o & -0.110(-0.119) & -0.086(-0.048) & 0.064(0.041) & 0.060(0.049) & -0.092(-0.098) & -0.043(0.011) \\

        \cellcolor{lightgray} & \cellcolor{lightgray}LLaMA2 & \cellcolor{lightgray}-0.079(0.008) & \cellcolor{lightgray}-0.203(-0.017) & \cellcolor{lightgray}-0.086(0.002) & \cellcolor{lightgray}\toptwohl{-0.218}(0.026) & \cellcolor{lightgray}-0.086(0.018) & \cellcolor{lightgray}\topthreehl{-0.210}(-0.011)	  \\
        
        \cellcolor{lightgray} & \cellcolor{lightgray}LLaMA3 & \cellcolor{lightgray} -0.106(-0.056) & \cellcolor{lightgray} \topthreehl{-0.159}(-0.032) & \cellcolor{lightgray} -0.062(-0.070) & \cellcolor{lightgray} -0.104(-0.091) & \cellcolor{lightgray} -0.093(-0.055) & \cellcolor{lightgray} -0.159(-0.032)	  \\

        \cellcolor{lightgray} & \cellcolor{lightgray}Gemma2 & \cellcolor{lightgray}-0.381(-0.172) & \cellcolor{lightgray}\toptwohl{-0.462}(-0.206) & \cellcolor{lightgray}-0.089(-0.004) & \cellcolor{lightgray}-0.158(-0.074) & \cellcolor{lightgray}\topthreehl{-0.398}(-0.179) & \cellcolor{lightgray}\toponehl{-0.519}(-0.212)	  \\

        \cellcolor{lightgray} & \cellcolor{lightgray}Phi3 & \cellcolor{lightgray}-0.208(-0.108) & \cellcolor{lightgray}\toptwohl{-0.231}(-0.058) & \cellcolor{lightgray}-0.120(-0.070) & \cellcolor{lightgray}-0.169(-0.073) & \cellcolor{lightgray}\toponehl{-0.235}(-0.122) & \cellcolor{lightgray}\topthreehl{-0.231}(-0.102)	  \\ 
        
        \cellcolor{lightgray} & \cellcolor{lightgray}GPT3 & \cellcolor{lightgray}0.168(0.179) & \cellcolor{lightgray}0.026(0.079) & \cellcolor{lightgray}\toptwohl{-0.236}(-0.246) & \cellcolor{lightgray}-0.066(-0.135) & \cellcolor{lightgray} 0.140(-0.003) & \cellcolor{lightgray}0.184(0.047) \\

        \cellcolor{lightgray} & \cellcolor{lightgray}GPT3.5 & \cellcolor{lightgray}0.003(-0.076) & \cellcolor{lightgray}0.065(0.049) & \cellcolor{lightgray}-0.087(-0.157) & \cellcolor{lightgray}0.072(0.056) & \cellcolor{lightgray}0.160(0.092) & \cellcolor{lightgray}0.183(0.125)\\

        \cellcolor{lightgray} & \cellcolor{lightgray}GPT4o mini & \cellcolor{lightgray}-0.109(-0.045) & \cellcolor{lightgray}\toponehl{-0.339}(-0.053) & \cellcolor{lightgray}-0.054(0.002) & \cellcolor{lightgray}-0.222(-0.150) & \cellcolor{lightgray}-0.131(-0.008) & \cellcolor{lightgray}-0.159(0.047) \\

        \multirow{-8}{*}{\cellcolor{lightgray}{Eli5-Category}} & \cellcolor{lightgray}GPT4o & \cellcolor{lightgray}0.115(0.108) & \cellcolor{lightgray}-0.017(0.144) & \cellcolor{lightgray}-0.158(-0.020) & \cellcolor{lightgray}-0.073(0.080) & \cellcolor{lightgray}-0.026(-0.036) & \cellcolor{lightgray}-0.001(-0.047) \\

        \multirow{8}{*}{Wiki-QA} & LLaMA2 & -0.057(0.050) & -0.081(0.014) & -0.064(-0.001) & -0.164(-0.036) & -0.077(0.023) & -0.111(-0.006) \\

        & LLaMA3 & -0.043(0.022) & -0.037(0.071) & -0.118(-0.114) & \topthreehl{-0.154}(-0.147) & -0.068(-0.014) & -0.032(0.081) \\

        & Gemma2 & -0.255(0.103) & \topthreehl{-0.480}(-0.034) & -0.094(0.301) & -0.233(0.163) & \toptwohl{-0.561}(-0.123) & \toponehl{-0.634}(-0.118) \\

        & Phi3 & -0.067(-0.001) & -0.132(0.002) & -0.152(-0.038) & \topthreehl{-0.238}(-0.033) & -0.181(-0.118) & -0.179(-0.083) \\
	
        & GPT3 & -0.064(0.049) & -0.204(-0.186) & -0.042(0.048) & \topthreehl{-0.247}(-0.146) & -0.042(-0.020) & -0.095(-0.111)	 \\
        
        & GPT3.5 & 0.092(0.065) & 0.078(0.056) & -0.017(-0.113) & -0.030(-0.109) & 0.157(0.014) & \topthreehl{0.214}(0.065) \\
        
        & GPT4o mini & \toponehl{-0.263}(\topthreehl{-0.196}) & -0.184(-0.116) & -0.004(0.053) & 0.014(0.008) & -0.150(-0.144) & -0.103(-0.110) \\

        & GPT4o & 0.023(-0.048) & \toptwohl{0.135}(0.075) & -0.000(0.027) & -0.071(-0.047) & -0.060(-0.076) & 0.081(0.055) \\
        
        \cellcolor{lightgray} & \cellcolor{lightgray}LLaMA2 & \cellcolor{lightgray}-0.022(-0.027)) & \cellcolor{lightgray}-0.036(0.058) & \cellcolor{lightgray}-0.064(-0.029) & \cellcolor{lightgray}\toponehl{-0.190}(0.024) & \cellcolor{lightgray}-0.026(-0.028) & \cellcolor{lightgray}-0.040(0.058)	 \\

        \cellcolor{lightgray} & \cellcolor{lightgray}LLaMA3 & \cellcolor{lightgray} 0.243(\toptwohl{0.369}) & \cellcolor{lightgray} 0.153(0.257) & \cellcolor{lightgray} -0.111(-0.090) & \cellcolor{lightgray} -0.335(-0.346) & \cellcolor{lightgray} 0.281(\toponehl{0.429}) & \cellcolor{lightgray} 0.293(\topthreehl{0.362})	 \\
        
        \cellcolor{lightgray} & \cellcolor{lightgray}Gemma2 & \cellcolor{lightgray}0.027(0.049) & \cellcolor{lightgray}0.243(0.100) & \cellcolor{lightgray}-0.016(0.048) & \cellcolor{lightgray}0.326(0.266) & \cellcolor{lightgray}0.052(0.027) & \cellcolor{lightgray}0.183(0.068) \\

        \cellcolor{lightgray} & \cellcolor{lightgray}Phi3 & \cellcolor{lightgray}-0.109(-0.102) & \cellcolor{lightgray}-0.060(-0.039) & \cellcolor{lightgray}-0.064(-0.088) & \cellcolor{lightgray}-0.033(-0.040) & \cellcolor{lightgray}-0.125(-0.123) & \cellcolor{lightgray}-0.052(-0.040) \\

        \cellcolor{lightgray} & \cellcolor{lightgray}GPT3 & \cellcolor{lightgray}-0.076(0.006) & \cellcolor{lightgray}0.016(0.021) & \cellcolor{lightgray}-0.033(0.095) & \cellcolor{lightgray}\topthreehl{0.215}(0.138) & \cellcolor{lightgray}-0.103(0.001) & \cellcolor{lightgray}0.013(0.040)  \\

        \cellcolor{lightgray} & \cellcolor{lightgray}GPT3.5 & \cellcolor{lightgray} -0.277(\toponehl{-0.363}) & \cellcolor{lightgray}-0.101(-0.081) & \cellcolor{lightgray}-0.099(-0.048) & \cellcolor{lightgray}-0.007(-0.003)  & \cellcolor{lightgray}-0.105(-0.201) & \cellcolor{lightgray}-0.031(-0.024)  \\

        \cellcolor{lightgray} & \cellcolor{lightgray}GPT4o mini & \cellcolor{lightgray} -0.001(-0.010) & \cellcolor{lightgray} -0.116(-0.030) & \cellcolor{lightgray} -0.091(-0.058) & \cellcolor{lightgray}  -0.133(-0.081) & \cellcolor{lightgray} -0.043(-0.034) & \cellcolor{lightgray} -0.064(-0.015)  \\

        \multirow{-8}{*}{\cellcolor{lightgray}WMT 2014} & \cellcolor{lightgray}GPT4o &  \cellcolor{lightgray} -0.003(0.018) &  \cellcolor{lightgray} 0.116(0.161) &  \cellcolor{lightgray} -0.058(-0.087) &  \cellcolor{lightgray} -0.062(-0.069) &  \cellcolor{lightgray} 0.060(-0.051) &  \cellcolor{lightgray} 0.102(0.080)  \\
        
        \bottomrule
    \end{tabular}
    \end{subtable}
    
\label{tab:evaluation_corrNLP}
\end{table*}

\vspace{1mm}

{\bf \noindent Experiment setups.} Due to the API constraints, an inference of closed-source LLMs (\ie, GPT3 and GPT3.5) could take up to 20 seconds. Consequently, evaluating an LLM on CNN/Daily Mail dataset with 11,490 instances would require nearly 575 hours. Therefore, we randomly sample each NLP task's dataset with 100 instances when evaluating closed-source LLMs. We further sample 40 out of 164 and 125 out of 500 instances for HumanEval and MBPP, respectively. {For closed-source models, we evaluated all instances.}


\vspace{1mm}
{\bf \noindent Hardware and software dependencies.} To conduct our large-scale experiments, we utilize a server with AMD 3955WX CPU (3.9GHz), 256GB RAM, and four NVIDIA A4000 GPUs (16GB VRAM of each). \responseline{The evaluation of open-source LLMs shown in the main text takes more than 2560 GPU hours, and the results on the website take around 864 GPU hours.}

\section{Results}
\label{sec:results}

\subsection{RQ1: Uncertainty Estimation for NLP Tasks}

To answer this research question, we evaluated \responseline{eight general LLMs}. We present the results of Pearson correlation coefficients between uncertainty scores and LLMs' performance in Table~\ref{tab:evaluation_corrNLP}. Such correlation is an indicator of whether the uncertainty estimation can predict LLMs' performance and further perform the risk assessment. The higher the absolute value of the coefficient, the stronger the correlation.
We investigate the results of NLP tasks from three perspectives:

\vspace{1mm}
{\noindent \textbf{Uncertainty Measurement Techniques.}} As can be observed from Table~\ref{tab:evaluation_corrNLP}, uncertainties estimated via \textit{sample-based} methods generally yield the highest correlation to an LLM's performance. 

\responseline{Specifically, \textit{sample-based} methods perform the best, achieving the highest correlation in 21 out of 32 scenarios across eight models and four tasks. Among these, Sample VRO leads in 12 scenarios and Sample VR in 9. In contrast, \textit{single-inference} methods only achieve one top position. This indicates that relying on single-inference uncertainty estimation might be unreliable in practical applications without a more refined strategy.}


\responseline{However, even for the most promising \textit{sample-based} methods, the majority correlation is below -0.6 (18 out of 21 cases) and their performance can still further space for enhancement (\eg, -1.00 correlation), calling for the design of more advanced techniques.}

\begin{tcbfinding}
\textbf{Finding~\thetcbcounter}: \textit{Sample-based VRO} achieves the best performance for \responseline{eight} different LLMs on NLP tasks in most cases, surpassing single-inference methods by a large margin. {While its potential is evident, further enhancement with advanced technique design is needed and promising for industrial deployment.}
\end{tcbfinding}

\textit{Perturbation-based} methods demonstrate moderate effectiveness, exceeding single-inference metrics in a substantial proportion of tasks. \responseline{It reaches the highest correlation in 10 out of 31 scenarios. For the selection of perturbation points, the maximum entropy point is usually better than the other two, counting for 50\% of the best cases.} Surprisingly, the minimum entropy point can sometimes work well for Llama2. 
We explored this issue further and found that Llama2 exhibits greater certainty compared to Llama1 and Llama3. For Llama2, we observed that it is substantially more ``certain'' than Llama-1. The 75th percentile of the entropy of all tokens generated by Llama2 is nearly 0, with a mean value of 0.18. Conversely, Llama-1 has a value of 2.62 at the 75th percentile, with a mean of 1.52. Llama-3 has a value of 0.391 at the 75th percentile, with a mean of 0.211. As such, the entropy selection strategy might not work well on Llama2. 

\responseline{Another interesting observation is that some models (\eg, LLaMA2, LlaMA3, Gemma2, Phi3, and also older models such as GPT-2 and LLaMA demonstrated on our website) exhibit high positive Pearson correlation coefficients of perturbation-based methods.}
The results indicate that, a higher uncertainty may even yield a better performance in such cases. Upon several case studies, we find that this observation could be attributed to the long input in the text summarization dataset. \responseline{These LLMs may lose focus on the input context, leading to a lack of understanding of the instructions for summarization. Instead, they mechanically continue the text immediately after the input, producing similar responses despite the stochasticity introduced by perturbations. This results in a low uncertainty score.} On the contrary, if an LLM understands the task description at the end of the prompt, it will respond to the perturbation more severely, causing a higher uncertainty.

\begin{tcbfinding}
{\textbf{Finding~\thetcbcounter}: Perturbation-based methods are more inclined to produce model and task-specific outcomes. Stakeholders might need to perform model-specific optimization if possible. }
\label{finding:model_specific}
\end{tcbfinding}
{\noindent \textbf{Influence of Distance Functions.}} We also find that task-specific distance does not provide better results compared with cosine distance between embeddings in both \textit{sample-based} and \textit{perturbation-based} methods. On average, using cosine distance achieves an increase of \responseline{0.034} for \textit{perturbation-based} methods and an increase of \responseline{0.137} for \textit{sample-based} methods across different tasks and LLMs.

\begin{tcbfinding}
\textbf{Finding~\thetcbcounter}: The cosine distance function yields a better performance on NLP tasks when leveraging stochasticity to estimate an LLM's prediction uncertainty.  
\end{tcbfinding}

\begin{figure}[t]
    \centering
    \includegraphics[width=0.99\linewidth]{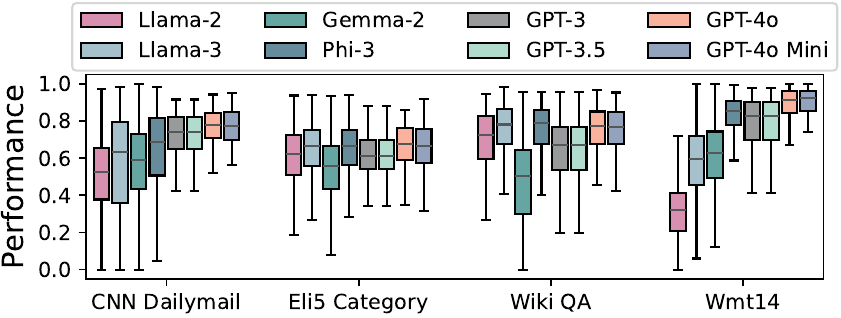}

    \caption{\responseline{Performance of LLMs on NLP tasks}}
    \label{fig:evaluation:nlp}
    \vspace{-4mm}
\end{figure}

{\noindent \textbf{Influence of Models.}} {Combining LLMs' performance in Figure~\ref{fig:evaluation:nlp} and uncertainty methods' effectiveness in Table~\ref{tab:evaluation_corrNLP}, we can observe that it is challenging for these uncertainty methods to assess the risks for models that are either \textit{exceptionally good} (\eg, GPT4o) or \textit{notably poor} (\eg, Llama2). This might be because the current evaluated methods struggle to discern the nuanced differences attributed to aleatoric uncertainty~\cite{hullermeier2021aleatoric} (knowledge deficiency). Most of the results are likely dominated by epistemic uncertainty~\cite{hullermeier2021aleatoric} (inherent randomness), leading to a decline in detection performance. In other words, when generated contents are unanimously good or bad, the computed uncertainty is more likely driven by inherent randomness rather than variations in the input data.}

\responseline{Another intriguing observation is that single-inference methods tend to perform better on more advanced models, such as the closed-source GPT series. This may be due to their extensive training data, resulting in better-calibrated confidence scores. However, a gap remains between single-inference methods and other approaches. Bridging this gap is challenging, as LLM output entropy consists of two components~\cite{arora2023theory}: language entropy, which arises from multiple vocabulary choices that convey similar meanings, and excess cross-entropy, which reflects model capability. Since single-inference methods rely on output entropy, separating these components is difficult. In contrast, multi-inference methods mitigate this issue by focusing on semantic variability rather than output entropy alone.}


%


\begin{tcbfinding}
\label{finding:nlp_subtle}
{\textbf{Finding~\thetcbcounter}: Uncertainty-based risk assessments for exceptionally good or notably poor LLMs appear to be limited. {The former is especially prevalent in closed-source commercial models used in real-world applications. When an LLM's performance is very high, analyzing its potential issues can be intrinsically difficult, calling for more advanced methods that can work even with limited information.}}
\end{tcbfinding}

\begin{table*}[htbp]
  \centering
  \caption{AUC scores for detecting erroneous code generated by LLMs. The results of VR/VRO are presented as ``{\em cosine distance-based (task-specific distance-based)}.'' Highest scores are ranked and highlighted as \colorbox{top1Color}{\textit{top-1}}, \colorbox{top2Color}{\textit{top-2}}, and \colorbox{top3Color}{\textit{top-3}}.}
  \renewcommand{\arraystretch}{1.1}
  \label{tab:evaluation_corrCode}
  \begin{subtable}[h]{\textwidth}
    \setlength{\tabcolsep}{11pt}
    \centering

    \scriptsize
        \resizebox{0.95\linewidth}{!}{
        \begin{tabular}{llccccccccc}
        \toprule
        \multirow{2}{*}{\textbf{Dataset}}&\multirow{2}{*}{\textbf{LLM}} & \multicolumn{4}{c}{{\bf Single-inference Method}} && \multicolumn{2}{c}{{\bf Sample-based Method}}\\
        \cmidrule{3-6} \cmidrule{8-9}
        && \textbf{Max Prob} & \textbf{Average Prob} & \textbf{Max Ent} & \textbf{Average Ent} && \textbf{Sample VR} & \textbf{Sample VRO}\\
        \cmidrule{1-6} \cmidrule{8-9}
        \multirow{7}{*}{{HumanEval}} & CodeQwen1.5 &0.592 & 0.539 & 0.578 & 0.538 && 0.639(0.617) & \toponehl{0.668}(\toptwohl{0.647})  \\

        & DeepSeek Coder & \topthreehl{0.735} & 0.692 & \toptwohl{0.738} & 0.676 && 0.730(0.705) & \toponehl{0.747}(0.699)  \\

        & Starcoder2 & \topthreehl{0.747} & 0.629 & \toponehl{0.752} & 0.656 && 0.693(0.675) & \toptwohl{0.748}(0.670)  \\

        & Code Llama& \topthreehl{0.744} & 0.618 & \toponehl{0.752} & 0.601 && 0.697(0.681) & 0.731(\toptwohl{0.749})\\

        & GPT3.5 & 0.500 & 0.590 & 0.383 & 0.490 && 0.630(\toponehl{0.667}) & \toptwohl{0.607}(0.587) \\

        & GPT4o mini & \toponehl{0.909} & \topthreehl{0.760} & \toptwohl{0.891} & 0.749 && 0.657(0.583) & 0.737(0.714) \\

        & GPT4o & 0.583 & 0.623 & 0.480 & 0.543 && 0.423(0.623) & 0.343(0.606) \\

        \cellcolor{lightgray}& \cellcolor{lightgray}CodeQwen1.5 & \cellcolor{lightgray}0.356 & \cellcolor{lightgray}0.363 & \cellcolor{lightgray}0.358 & \cellcolor{lightgray}0.367 && \cellcolor{lightgray}0.555(\toptwohl{0.613}) & \cellcolor{lightgray}\topthreehl{0.593}(\toponehl{0.643})   \\
        
        \cellcolor{lightgray}& \cellcolor{lightgray}DeepSeek Coder & \cellcolor{lightgray}0.545 & \cellcolor{lightgray}0.463 & \cellcolor{lightgray}0.527 & \cellcolor{lightgray}0.46 && \cellcolor{lightgray}0.581({0.618}) & \cellcolor{lightgray}0.617(\topthreehl{0.621})	  \\

        \cellcolor{lightgray}& \cellcolor{lightgray}Starcoder2 & \cellcolor{lightgray} 0.567 & \cellcolor{lightgray} 0.447 & \cellcolor{lightgray} 0.589 & \cellcolor{lightgray} 0.439 && \cellcolor{lightgray} 0.562(0.616) & \cellcolor{lightgray} 0.581(\toponehl{0.665})  \\

        \cellcolor{lightgray}& \cellcolor{lightgray}Code Llama & \cellcolor{lightgray}0.499 & \cellcolor{lightgray}0.418 & \cellcolor{lightgray}0.483 & \cellcolor{lightgray}0.433 && \cellcolor{lightgray}0.578(\topthreehl{0.668}) & \cellcolor{lightgray}0.603(\toponehl{0.705})\\

        \cellcolor{lightgray} & \cellcolor{lightgray}GPT3.5 & \cellcolor{lightgray}\topthreehl{0.582} & \cellcolor{lightgray}0.556 & \cellcolor{lightgray}\toptwohl{0.606} & \cellcolor{lightgray}0.560 && \cellcolor{lightgray}0.592(0.563) & \cellcolor{lightgray}0.561(0.576)  \\

         \cellcolor{lightgray}& \cellcolor{lightgray}GPT4o mini & \cellcolor{lightgray} 0.579 & \cellcolor{lightgray} 0.609 & \cellcolor{lightgray} 0.593 & \cellcolor{lightgray} \topthreehl{0.628} && \cellcolor{lightgray} 0.609(0.606) & \cellcolor{lightgray} 0.599(0.569)  \\

        \multirow{-7}{*}{\cellcolor{lightgray}{MBPP}} & \cellcolor{lightgray}GPT4o & \cellcolor{lightgray} 0.629 & \cellcolor{lightgray} 0.633 & \cellcolor{lightgray} 0.635 & \cellcolor{lightgray} \topthreehl{0.667} && \cellcolor{lightgray} 0.650(\toponehl{0.735}) & \cellcolor{lightgray} 0.634(\toptwohl{0.698})  \\

        \bottomrule
        \end{tabular}
        }
  \end{subtable}
  \vspace{0.5mm}
  \vfill
  \begin{subtable}[h]{\textwidth}
  \setlength{\tabcolsep}{10.28pt}
  \centering

  \scriptsize
  \resizebox{0.95\linewidth}{!}{
    \begin{tabular}{llcccccc}
        \toprule
        \multirow{2}{*}{\textbf{Dataset}}&\multirow{2}{*}{\textbf{LLM}} & \multicolumn{6}{c}{\textbf{Perturbation-based}}\\
        \cmidrule{3-8}
        && \textbf{Max VR} & \textbf{Max VRO} & \textbf{Min VR} & \textbf{Min VRO} & \textbf{MaxDiff VR} & \textbf{MaxDiff VRO} \\
        \midrule
        
        \multirow{7}{*}{HumanEval} & CodeQwen1.5 & 0.629(0.611) & \topthreehl{0.646}(0.609) & 0.595(0.592) & 0.593(0.571) & 0.626(0.627) & 0.638(0.595)
         \\

        & DeepSeek Coder & 0.607(0.479) & 0.664(0.512) & 0.589(0.563) & 0.574(0.426) & 0.604(0.454) & 0.661(0.581)	  \\

         & Starcoder2 & 0.638(0.569) & 0.664(0.643) & 0.449(0.457) & 0.565(0.526) & 0.592(0.525) & 0.666(0.629) \\

         & Code Llama & 0.597(0.613) & 0.668(0.644) & 0.498(0.555) & 0.643(0.582) & 0.609(0.538) & 0.725(0.680)  \\

        & GPT3.5 & 0.500(0.503) & 0.457(0.560) & 0.497(0.503) & 0.420(\topthreehl{0.607}) & 0.533(0.533) & 0.393(0.477)  \\

        & GPT4o mini & 0.546(0.531) & 0.703(0.743) & 0.594(0.531) & 0.589(0.646) & 0.506(0.460) & 0.686(0.749)  \\

        & GPT4o & 0.571(0.560) & \toptwohl{0.709}(\toponehl{0.720}) & 0.391(0.469) & 0.600(0.674) & 0.457(0.509) & \topthreehl{0.691}(0.680) \\

        \cellcolor{lightgray} & \cellcolor{lightgray}CodeQwen1.5 & \cellcolor{lightgray}0.533(0.532) & \cellcolor{lightgray}0.501(0.553) & \cellcolor{lightgray}0.476(0.478) & \cellcolor{lightgray}0.487(0.520) & \cellcolor{lightgray}0.567(0.553) & \cellcolor{lightgray}0.523(0.558) \\

        \cellcolor{lightgray} & \cellcolor{lightgray}DeepSeek Coder & \cellcolor{lightgray}0.620(0.592) & \cellcolor{lightgray}0.561(0.566) & \cellcolor{lightgray}0.456(0.480) & \cellcolor{lightgray}0.459(0.487) & \cellcolor{lightgray}\toponehl{0.658}(\toptwohl{0.654}) & \cellcolor{lightgray}0.619(0.619)  \\

        \cellcolor{lightgray} & \cellcolor{lightgray}Starcoder2 & \cellcolor{lightgray} 0.600(0.576) & \cellcolor{lightgray} 0.568(0.587) & \cellcolor{lightgray} 0.485(0.451) & \cellcolor{lightgray} 0.420(0.477) & \cellcolor{lightgray} \toptwohl{0.635}(\topthreehl{0.632}) & \cellcolor{lightgray} 0.570(0.627) \\

        \cellcolor{lightgray} & \cellcolor{lightgray}Code Llama & \cellcolor{lightgray}0.586(0.630) & \cellcolor{lightgray}0.521(0.621) & \cellcolor{lightgray}0.454(0.507) & \cellcolor{lightgray}0.501(0.568) & \cellcolor{lightgray}0.622(\toptwohl{0.679}) & \cellcolor{lightgray}0.566(0.667)	 \\
        
        \cellcolor{lightgray} & \cellcolor{lightgray}GPT3.5 & \cellcolor{lightgray}0.587(\toponehl{0.628}) & \cellcolor{lightgray}0.511(0.544) & \cellcolor{lightgray}0.550(0.495) & \cellcolor{lightgray}0.508(0.511) & \cellcolor{lightgray}0.498(0.571) & \cellcolor{lightgray}0.486(0.528)  \\

        \cellcolor{lightgray} & \cellcolor{lightgray}GPT4o mini & \cellcolor{lightgray} 0.574(0.487) & \cellcolor{lightgray} 0.574(0.463) & \cellcolor{lightgray} 0.416(0.397) & \cellcolor{lightgray} 0.420(0.435) & \cellcolor{lightgray} \toptwohl{0.650}(0.551) & \cellcolor{lightgray} \toponehl{0.675}(0.533)	 \\

        \multirow{-7}{*}{\cellcolor{lightgray}MBPP} & \cellcolor{lightgray}GPT4o & \cellcolor{lightgray}0.448(0.538) & \cellcolor{lightgray}0.418(0.546) & \cellcolor{lightgray}0.560(0.591) & \cellcolor{lightgray}0.536(0.549) & \cellcolor{lightgray}0.514(0.545) & \cellcolor{lightgray}0.454(0.495)	 \\
        
        \bottomrule
    \end{tabular}
    }
    \end{subtable}
\end{table*}

\begin{table*}[htbp]
  \centering
  \caption{Pearson correlation coefficients of LLaMA2 without system prompts on four NLP tasks. The results of VR/VRO are presented as ``{\em cosine distance-based (task-specific distance-based)}.'' Highest correlations from different categories are ranked and highlighted as \colorbox{top1Color}{\textit{top-1}}, \colorbox{top2Color}{\textit{top-2}}, and \colorbox{top3Color}{\textit{top-3}}. 
  }
  \renewcommand{\arraystretch}{1.1}
  \begin{subtable}[h]{\textwidth}
    \setlength{\tabcolsep}{12pt}
    \centering
    \scriptsize
        \begin{tabular}{lcccccll}
        \toprule
        \multirow{2}{*}{\textbf{Dataset}} & \multicolumn{4}{c}{{\bf Single-inference Method}} && \multicolumn{2}{c}{{\bf Sample-based Method}}\\
        \cmidrule{2-6} \cmidrule{7-8}
        & \textbf{Max Prob} & \textbf{Average Prob} & \textbf{Max Ent} & \textbf{Average Ent} && \textbf{Sample VR} & \textbf{Sample VRO}\\
        \cmidrule{1-6} \cmidrule{7-8}
        CNN/Daily Mail   & -0.349 & -0.480 & -0.435 & -0.528 && -\toptwohl{0.601}(\topthreehl{-0.528}) & \toponehl{-0.638} (-0.522)  \\
        Eli5-Category & -0.054 & {-0.316} & -0.182 & {-0.362}  && -0.354(-0.020) & \topthreehl{-0.444}(0.006)  \\
        Wiki-QA & -0.093 & -0.122 & -0.073 & -0.173  && \toptwohl{-0.534}(-0.212) & \toponehl{-0.589}(-0.138)  \\
        WMT 2014 & {-0.006} & 0.176 & -0.145 & -0.003 && \toptwohl{-0.611}(-0.527) & \toponehl{-0.746}(-0.578)  \\
        
        \bottomrule
        \end{tabular}
  \end{subtable}
  \vspace{6pt}
  \vfill
  \begin{subtable}[h]{\textwidth}
  \setlength{\tabcolsep}{11.85pt}
  \centering
  \scriptsize
    \begin{tabular}{lcccccc}
        \toprule
        \multirow{2}{*}{\textbf{Dataset}} & \multicolumn{6}{c}{\textbf{Perturbation-based}}\\
        \cmidrule{2-7}
        & \textbf{Max VR} & \textbf{Max VRO} & \textbf{Min VR} & \textbf{Min VRO} & \textbf{MaxDiff VR} & \textbf{MaxDiff VRO} \\
        \midrule
        
        CNN/Daily Mail & 0.098(-0.054) & -0.006(-0.053)	 & 0.000(-0.087) & -0.332(-0.320) & 0.123(-0.037) & 0.025(-0.028) \\

        Eli5-Category & -0.191(-0.124) & \toponehl{-0.513}(-0.293)	 & -0.183(-0.042)	 & -0.256(0.059) & -0.169(-0.109) & \toptwohl{-0.498}(-0.276)  \\
        
        Wiki-QA &  0.007(0.029) & -0.008(0.070)	 & -0.193(-0.123) & \topthreehl{-0.438}(-0.103) & 0.049(0.010) & -0.058(0.011)  \\
       
        WMT 2014  &  0.034(-0.066) & 0.230(-0.032) & -0.400(-0.409)	& \topthreehl{-0.608}(-0.515) & 0.030(-0.055) & 0.266(0.014) \\
    
        \bottomrule
    \end{tabular}
    \end{subtable}
    
\label{tab:llama2_extra}
\end{table*}


\subsection{RQ2: Limitations in Uncertainty Estimation for NLP Tasks}

{In this RQ, we further explore two limitations of evaluated uncertainty estimation methods.}


{
{\noindent \textbf{Influence of prompt.}} An intriguing anomaly in RQ1 is Llama2, which underperforms significantly on summarization and translation, and its uncertainty estimation for all four tasks is ineffective. This is counter-intuitive. After manually investigating some randomly chosen samples, we found this might be because of the input prompt. Using a fine-tuned chat version of Llama2, we incorporated a system prompt template shown in the original paper, starting with 

\begin{tcolorbox}[colframe=white,colback=white, top=0pt, bottom=0pt, before=\vspace{5pt}, after=\vspace{5pt}, left=0pt,]
{\ttfamily You are a helpful, respectful, and honest assistant ...}
\end{tcolorbox}

However, this prompt template has three downsides. We will use the WMT-14 translation task as examples: (1)  The LLM will sometimes directly reject to respond, offering something such as `` I cannot provide a translation for that statement as it is not factually coherent.'' (2) The LLM misinterprets examples in the in-context prompt, translating the (fixed) example itself. This leads to duplicate translations and results in poor translation performance. (3) Many of the responses will begin with greetings such as  `` Thank you for your kind and respectful instructions! '' Each of these three scenarios can negatively affect the calculation of multi-inference uncertainty, as the fine-tuned responses from the LLM, to some extent, surpass inherent randomness. 

To study the prompt influence further, we removed the system prompt and re-run all the experiments on Llama2. The results are shown in Table~\ref{tab:llama2_extra}. Overall, there's a significant increase in all correlations, encompassing the original two high-performing QA tasks. When comparing the top-1 scores across the four tasks, the average increase in the absolute value of correlation coefficients is 0.272. The most notable improvement is observed in the WMT-14 dataset, where the correlation shifts from -0.190 to -0.746. {Given that RLHF-based prompt tuning is pivotal for SOTA chat models, such as open-sourced Llama2 and commercial ChatGPT, it might be better for future work to take this into consideration when designing new methods.}

\begin{tcbfinding}
{ 
    \textbf{Finding~\thetcbcounter}: Prompts could significantly influence uncertainty estimation in some cases. Prompt templates can lead LLMs to exhibit different behaviors, compromising the accuracy of uncertainty estimation.
}
\end{tcbfinding}



{\noindent \textbf{Relation of uncertainty and inaccuracy.}} We find that a low uncertainty does not guarantee that an LLM's response is reliable. 
Specifically, an LLM can generate highly confident responses while with non-factual information. 


\responseline{A representative example can be observed when Gemma2 is instructed to summarize a CNN news article in which a triathlon participant narrates her training journey in the first person.}

\begin{tcolorbox}[colframe=white,colback=white, top=0pt, bottom=0pt, before=\vspace{2pt}, after=\vspace{2pt}, left=0pt,]
\responseline{\ttfamily 
... I'm looking forward to many more rides outdoors. I want to say thank you to the fellowship of fitness that I've been lucky enough to find ... I'm so grateful that ...
}
\end{tcolorbox}



\noindent \responseline{The article contains many first-person narratives, which cause the model to misinterpret the task and ignore the instruction to summarize. As a result, it repeated sentences such as ``{\ttfamily I'm still not doing as much as xxx as I should ...}'' In the response, the LLM consistently produced similar responses, even when perturbation or sampling was introduced to add stochasticity. Despite the fact that these responses were entirely incorrect, the LLM still generated a low uncertainty score. We provide the full example on our website.}

Conversely, a higher degree of uncertainty also does not necessarily imply that an LLM's prediction is incorrect. We show an example of GPT 3.5, which summarizes a piece of news from the CNN/Daily Mail dataset. 
In this case, the ground truth summary includes two details: \textit{a fire occurrence at a park} and \textit{the absence of injuries}. 
For sample-based uncertainty measurement, all five generated samples incorporated these two pieces of information but also furnished additional varied information, such as the park's owner and its intended use. Such extra information further leads to a higher variance in the generated samples' embedding, resulting in a high uncertainty score despite that the LLM's prediction is reliable.

\begin{tcbfinding}

{\textbf{Finding~\thetcbcounter}: Uncertainty is not always correlated to inaccuracy. {Future work may also consider combining other features or indicators (besides uncertainty) to the risk assessment for better performance.}}
\end{tcbfinding}

\subsection{RQ3: Uncertainty Estimation for Code Generation}

{To answer this research question, we evaluate \responseline{seven} LLMs capable of code generation. Their performances are shown in Figure~\ref{fig:evaluation:code}. When evaluating the efficacy of uncertainty methods in identifying erroneous code, we treat the problem as a binary detection task. Code is labeled as 1 if completely correct (\eg, $Q=1$) and 0 if not. We take the uncertainty scores as indicators, and higher uncertainty suggests a greater likelihood of errors in the code. Subsequently, we calculate the Area Under the Receiver Operating Characteristic Curve (AUC) scores for each method. The results are shown in Table~\ref{tab:evaluation_corrCode}.}



{\noindent \textbf{Uncertainty Measurement Techniques.}}
\responseline{Even with the distinct task of code generation and the different evaluation metrics of AUC, we can still observe a similar trend on code generation tasks than on NLP tasks, \ie, \textit{sample-based} methods dominate across two different datasets and diverse LLMs. Perturbation-based methods remain in second place, with single-inference methods ranking last overall. However, a nuanced distinction is that single-inference methods perform better in code generation tasks compared to other NLP tasks. Specifically, they achieved 3 top-1 AUC scores out of 14 cases (21.4\%), while for NLP tasks, they only secured one top-1 correlation out of 32 cases (3.13\%). One possible explanation is that programming languages are more deterministic than natural languages, which reduces the impact of language entropy~\cite{arora2023theory} on uncertainty estimation.}



\begin{tcbfinding}
\label{Finding:sample_in_code}
{\textbf{Finding~\thetcbcounter}: Sample-based methods remain the most effective for risk assessment in code generation. Although single-inference methods perform worse than perturbation-based methods, they show stronger performance on code generation tasks compared to NLP tasks.}
\end{tcbfinding}

{\noindent \textbf{Influence of Distance Functions.}} Different from the results on NLP tasks, the cosine distance between code embeddings under-performs significantly in comparison with the task-specific distance function, CodeBLEU. \responseline{Task-specific metric prevails in 14 out of 24 cases for the TOP-3 results. }


\begin{tcbfinding}
\label{Finding:codebleu}
{\textbf{Finding~\thetcbcounter}: 
The distance function can play an important role in uncertainty estimation as indicated by the result difference between domains of NLP and code.
A more carefully designed distance function that suits downstream tasks could potentially enhance the effectiveness of risk assessment.}
\end{tcbfinding}

\begin{figure}[t]
    \centering
    \includegraphics[width=0.99\linewidth]{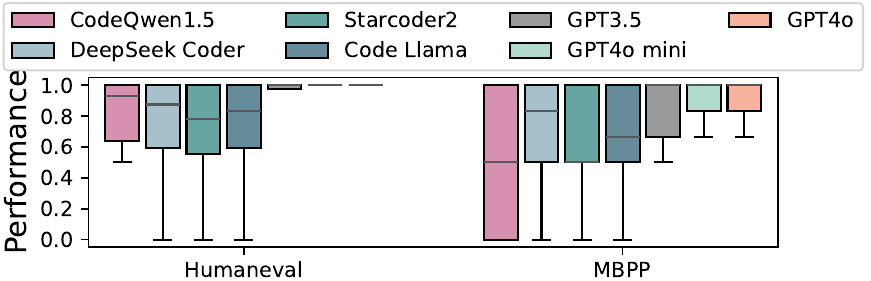}
    \caption{{\responseline{$Q$ scores of LLMs on code generation tasks}}}
    \label{fig:evaluation:code}
    \vspace{-15pt}
\end{figure}

{\noindent \textbf{Influence of Models.}} \responseline{Among all models, GPT-3.5 is the most challenging model to assess code quality using uncertainty estimation. The best AUC score is 0.667 on the HumanEval dataset and 0.628 on the MBPP dataset. Both of them are below 0.7. According to Figure~\ref{fig:evaluation:code}, GPT-3.5 demonstrates relatively strong performance compared to most models but still falls short of GPT4o and GPT4o mini. It is expected that GPT-3.5 has a lower AUC score than other open-source models because, as seen in our findings on NLP tasks, nuanced errors are more difficult to detect using uncertainty metrics. The errors made by GPT-3.5 tend to be subtler than those made by open-source models. However, it may seem counterintuitive that assessing risks for GPT-3.5 is more challenging compared to GPT4o and GPT4o mini. Upon further exploration, we found that GPT-3.5 is more prone to making \textit{partial errors}. For instance, on the MBPP dataset, 40.4\% of GPT-3.5’s errors passed at least one test case, compared to 27.3\% for GPT4o and 32.4\% for GPT4o mini. This suggests that the GPT4 family is more likely to either fully misunderstand the question or generate a completely correct response, making their errors easier to detect.
}




\subsection{RQ4: Limitations in Uncertainty Estimation for Code Generation}

{\noindent \textbf{Limitations of detecting subtle errors.}} In RQ3, we hypothesized that the relatively poor performance of uncertainty estimation in the detection of GPT-3.5's errors is due to the challenges in pinpointing subtle mistakes that existed in the generated code. To validate this hypothesis, we conducted additional experiments in this RQ. \responseline{We evaluated all the methods across seven LLMs (with results for three available on our website)} on two datasets under two settings. In setting \textit{A}, only syntactically correct code is kept (\eg, $Q \geq 0.5$). In setting \textit{B}, only syntactically incorrect and completely correct code is kept (\eg, $Q=1$ or $Q < 0.5$). We excluded the GPT family since all their code is syntactically correct. We recorded the best AUC score for each model under both settings. We leave the full result on our website.

\responseline{Comparing the performances between setting \textit{A} and setting \textit{B}, we observe that, with the exception of Santacoder on HumanEval, in which we exclude details in the main content due to page limit,
all other LLMs show a notable improvement. The average score increase is 0.095.} 
Santacoder's decline might be attributed to its frequent generation of syntactically incorrect code, making the partially correct code an easily detectable outlier.

\begin{tcbfinding}
\label{finding:suble}
    {\textbf{Finding~\thetcbcounter: } Current uncertainty methods are more suitable for detecting obvious errors instead of subtle errors.}
\end{tcbfinding}




\vspace{1mm}
\noindent \textbf{Limitations of distance functions.} Different from our observations in RQ1, the cosine distance metric does not show dominant performance compared with task-specific metrics on estimating LLMs' risks when generating code. This indicates that it is non-trivial for selected embedding model~\cite{guo2020graphcodebert} to detect minor code differences due to an LLM's stochasticity.

The efficiency of multi-inference metrics heavily depends on the precise estimation of the distance between data points. Both RQ1 and RQ3 highlight the crucial role of the distance function. However, it appears that the cosine distance metric falls short in code-related tasks. It is non-trivial for current embedding LLMs to detect subtle differences generated by randomness or perturbations. 

We calculated pairwise distances within each group of responses generated via sample-based and perturbation-based methods. As shown in Table~\ref{tab:code_dist}, the average cosine distance is significantly lower than its task-specific counterpart (\ie, CodeBLEU~\cite{ren2020codebleu}). This directly affects the effectiveness of \textit{VR}/\textit{VRO} when determining the degree of uncertainty. 

Nevertheless, CodeBLEU~\cite{ren2020codebleu} also comes with its limitations. For instance, two programs that only differ from variable names might lead to a large CodeBLEU distance, resulting in erroneous uncertainty estimation. 

\begin{tcbfinding}
    {\textbf{Finding~\thetcbcounter: } 
    Neither cosine distance nor CodeBLEU could accurately assess the difference between the two programs, resulting in under-performed uncertainty estimation for LLMs compared with NLP tasks.
    Characterizing the true difference between code can be challenging and is a limitation to performing accurate uncertainty estimation. }
\end{tcbfinding}

\begin{table}[!tb]
    \caption{The average distances between generated code.}
    \centering
    \responseref{}
    \renewcommand{\arraystretch}{1.1}
    \setlength{\tabcolsep}{6pt}
    \scriptsize
    \label{tab:code_dist}
        \setlength{\tabcolsep}{3pt}
        \begin{tabular}{lccccc}
        \toprule
        & \multicolumn{2}{c}{\textbf{Perturbation}} &&  \multicolumn{2}{c}{\textbf{Sample}}  \\
        \cmidrule{2-3} \cmidrule{5-6}
        & Open-source & Closed-source  && Open-source & Closed-source \\
        \midrule
        Cosine & 0.030$\pm$0.034 & 0.015$\pm$0.036  && 0.027$\pm$0.017 & 0.009$\pm$0.010 \\
        CodeBLEU & 0.513$\pm$0.198 & 0.260$\pm$0.213  && 0.679$\pm$0.111 & 0.374$\pm$0.210 \\
        \bottomrule
        \end{tabular}
\end{table}

\section{Implication and Opportunity}
\label{sec:discussion}

\subsection{\responseline{For Researchers}}

\noindent \textbf{Prompt is important.} Prompt has long been proven to be a key factor in LLM's performance. In our study, we further demonstrate its potential to significantly affect uncertainty estimation's efficacy (Finding 5). Specifically, the uncertain behavior of LLMs might be profoundly impacted by the prompt used in RLHF~\cite{ouyang2022training}. Original uncertain answers may be supplanted by human-favored responses when integrating specific prompt templates. A prospective research direction is to explore the influence of the RLHF process on uncertainty estimation and to discern strategies for more accurate estimations, both from training (\eg, calibrate the model better) and inference (\eg, refine estimation methods) standpoints. Another interesting perspective is to design a better prompt (\eg, instruct LLMs to switch to the uncertainty estimation mode) to enhance the precision of the measurement.


\vspace{1mm}
\noindent \textbf{Subtle errors can be hard to detect.} We observed that selected methods could struggle to detect subtle errors in partially correct code and are easier to obvious errors (Finding 9). One possible future direction is to improve their sensitivity by separating the uncertainty caused by model inability and that stemming from inherent randomness. Another direction could involve constructing a multi-stage system, with uncertainty-related methods at the forefront, followed by other techniques (\eg, white-box).

\vspace{1mm}

\noindent \textbf{Better perturbation strategy is needed for more accurate uncertainty estimation.} Our \textit{perturbation-based} methods leverage the unique characteristic of an autoregressive language model to perturb its decoding process, which shows moderate uncertainty estimation performance in the experiments. Compared with \textit{sample-based} methods, the \textit{perturbation-based} methods do not require access and tuning the temperature setting ($T$). Despite the fact that the \textit{perturbation-based} methods underperform the \textit{sample-based} methods in general, we believe the \textit{perturbation-based} methods could be further improved with, \eg, a more fine-grained strategy to identify key point for perturbation.

\vspace{1mm}

\noindent \responseline{\textbf{Using uncertainty estimation alone might not be enough for comprehensive risk assessment.} In RQ2 and Finding 6, we discussed and concluded that a model's uncertainty does not necessarily indicate its correctness. LLMs can be highly confident in incorrect answers and vice versa. Therefore, relying solely on uncertainty measurement may not be sufficient for practical LLM deployment. A promising direction is to incorporate behavioral testing~\cite{ribeiro-etal-2020-beyond, ferrando-etal-2023-automating, lee2024automated}, a method recently proposed for evaluating the correctness of NLP systems. Rather than assessing whether LLMs are fully accurate, behavioral testing evaluates whether their output meets certain important properties. This more flexible approach could potentially be automated without the need for ground truth and might help filter out clearly unreasonable responses, such as those that repeatedly generate meaningless sentences. We believe that integrating behavioral testing with uncertainty measurement into a comprehensive risk assessment framework will open new opportunities for improving the reliability of LLM-driven systems.}

\subsection{\responseline{For Developers}}

\noindent \textbf{Ask more, get more.} In our study, \textit{multi-inference} methods perform better than single-inference methods across different tasks in most cases. As for a \textit{black-box} LLM, getting a comprehensive understanding beforehand or only through a single deterministic inference could be challenging. Instead, the more we query an LLM, the clearer we can get about its internal knowledge regarding a specific aspect.
We hypothesize that this is because by querying models multiple times, we gain more knowledge from them. This might be a promising technique when the model is black-box. 

\vspace{1mm}

\noindent \textbf{Model-specific uncertainty estimation might be beneficial.} Although all the methods chosen in the paper are black-box, we observed considerable variations in their effectiveness across different models (\eg, Finding 2). Thus, to enhance the efficacy of an uncertainty-based risk assessment system, stakeholders might need to tailor their methods and undertake model-specific optimizations. These adjustments could be necessary even between different model versions (\eg, LlaMA2 and LlaMA3). \responseline{Another important factor that could influence the effectiveness of uncertainty measurement is the deployment choices, such as quantization types and even the backend hardware. These factors may affect computational accuracy and create subtle variations between the original models. In practical scenarios, models can be versioned based on their quantization level, which might significantly impact their performance and computational efficiency~\cite{li2024evaluating}. We believe this is an intriguing direction for future exploration and will leave it for future work.}


\section{Threats to Validity}
\label{sec:threats}


In terms of \textbf{internal threats}, the selection of uncertainty estimation techniques can be a threat that affects the validity of our findings.
In our study, we tried our best and collect as many as 12 existing uncertainty estimation methods from different categories (single deterministic, Bayesian, and test-time augmentation), to better understand the effectiveness of uncertainty estimation under the scope of LLMs' erroneous generations.

In terms of \textbf{external threats}, the subject LLM selection for the evaluation could be another threat. \responseline{The generalizability of our uncertainty estimation methods to LLMs beyond those studied presents a potential threat, as LLMs may exhibit different uncertainty characteristics across various tasks. Additionally, our findings and conclusions may differ as LLM technology advances. To address this, we selected a diverse range of 17 LLMs from both open-source and closed-source candidates, spanning from GPT-2 XL (released in 2019) to GPT4o (released very recently). We have carefully reviewed our conclusions and made every effort to present findings that are relevant and useful for state-of-the-art LLMs. Still, future models may exhibit different characteristics. To support continued research, we will open-source our codebase, allowing future researchers to test and analyze emerging models as they are developed.}

\responseline{One limitation of this work is that we focus solely on evaluating code generation for code-specific LLMs. Other tasks, such as program repair~\cite{xia2023automated} and code translation~\cite{translation2024}, also merit consideration. We leave a more in-depth analysis of these code-related tasks for future research.}



\section{Conclusion}
\label{sec:conclusion}
This paper initiates an early exploratory study toward understanding the risk assessment of LLMs from the lens of uncertainty estimation. 
A large-scale evaluation of twelve different uncertainty estimation techniques on nine LLMs and six tasks is conducted.
A further in-depth analysis was made to investigate the correlations between LLMs' prediction uncertainty and their performance.
Understanding the potential risks of LLMs could be of great importance for industrial-scale applications. 
Our results confirm that uncertainty estimation can be a promising direction for potential risk assessment of LLMs in both NLP and code-generation domain tasks. 
However, there can still be much space and opportunity to design more advanced uncertainty estimation techniques to characterize the risks of an LLM more effectively. 
Moreover, other possibly useful quality indicators besides uncertainty could also be designed to better characterize the capability boundary of an LLM from multiple perspectives. 
With the recently increasing demand and urgency for trustworthiness assurance of LLMs in industry, we hope this paper could potentially inspire researchers and practitioners, to join the force to design novel techniques and toolchain support and together conquer many new relevant challenges. We also make the replication package of this paper available, to enable further research towards realizing trustworthy LLMs for industrial usage. 






\section*{Acknowledgments}


This work was supported in part by JST CRONOS Grant (No.JPMJCS24K8), JST-Mirai Program Grant (No.JPMJMI20B8), JSPS KAKENHI Grant (No.JP21H04877, No.JP23H03372, and No.JP24K02920), Canada CIFAR AI Chairs Program, the Natural Sciences and Engineering Research Council of Canada, and the Autoware Foundation.

\bibliographystyle{IEEEtran}
\bibliography{reference}

\begin{IEEEbiography}[{\includegraphics[width=1in,height=1.25in,clip,keepaspectratio]{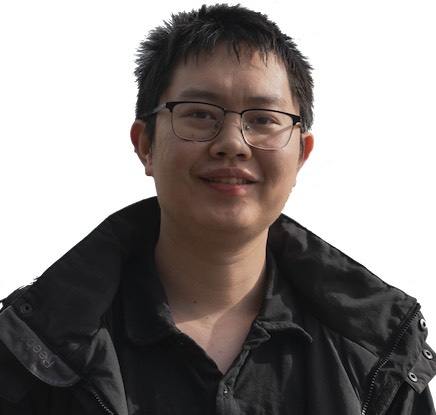}}]{Yuheng Huang}(Graduate Student Member, IEEE)
received a BE degree from the Beijing University of Posts and Telecommunications, China, and a master’s degree from the University of Alberta, Canada. He is currently working toward a PhD degree at The University of Tokyo, Japan. His research interests primarily focus on the quality assurance of AI-enabled complex systems as well as developing interactive interfaces that can enable more efficient human-in-the-loop AI development from Human-Computer Interaction (HCI) perspectives.
\end{IEEEbiography}

\vspace{-10mm}

\begin{IEEEbiography}[{\includegraphics[width=1in,height=1.25in,clip,keepaspectratio]{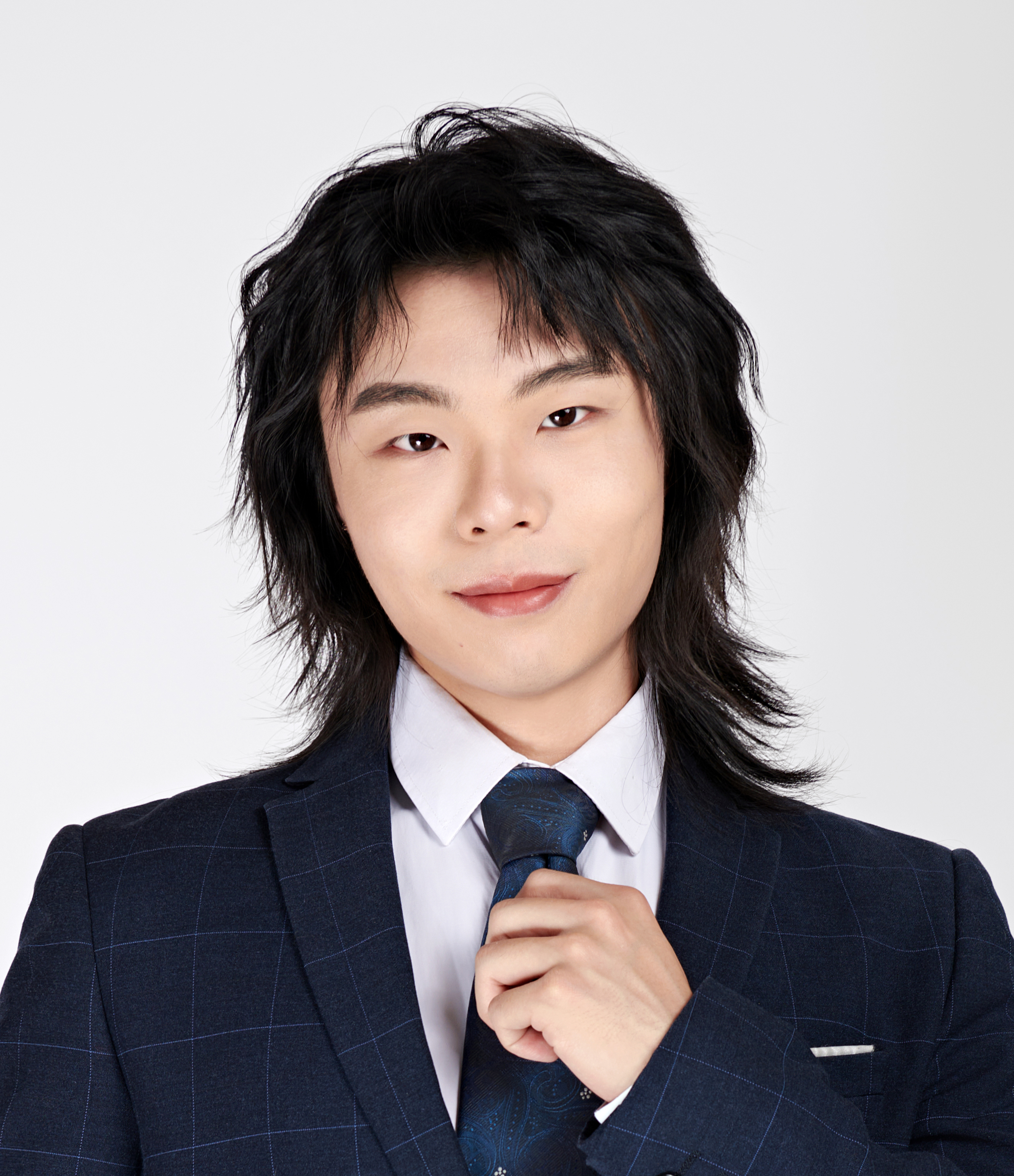}}]{Jiayang Song}
(Graduate Student Member, IEEE)
received the B.E. degree from Western University, London, ON, Canada, in 2019 and the M.E. degree from the University of Toronto, Toronto, ON, Canada, in 2021. He is currently pursuing a Ph.D. degree with the University of Alberta, Edmonton, AB, Canada.
His research interests include testing, analysis, repairing, and enhancement of AI systems and their applications for quality assurance of trustworthy AI-enabled cyber-physical systems.
\end{IEEEbiography}

\vspace{-10mm}

\begin{IEEEbiography}
[{\includegraphics[width=1in,height=1.25in,clip]{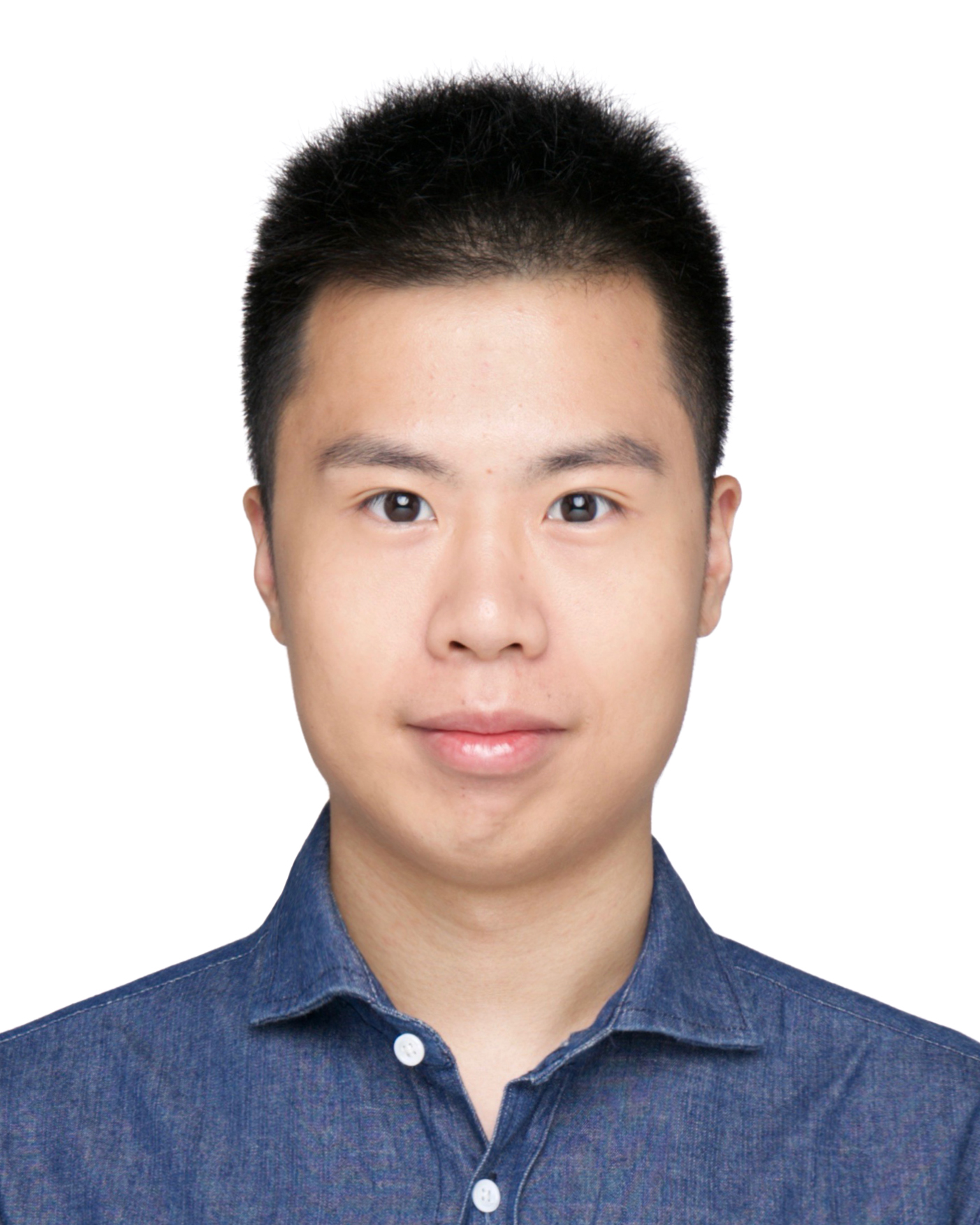}}]
{Zhijie Wang}
(Graduate Student Member, IEEE)
is currently a Ph.D. candidate in Software Engineering at the University of Alberta, AB, Canada. Previously, he received his M.E. degree from the University of Waterloo, ON, Canada in 2021. His research interest focuses on software engineering support for complex AI-based software systems. He is also broadly interested in the intersection of software engineering, AI, and human-computer interaction (HCI). His work has been published in top-tier SE and HCI venues and has received a best paper award (FSE '23).
\end{IEEEbiography}

\vspace{-10mm}

\begin{IEEEbiography}
[{\includegraphics[width=1in,height=1.25in,clip,keepaspectratio]{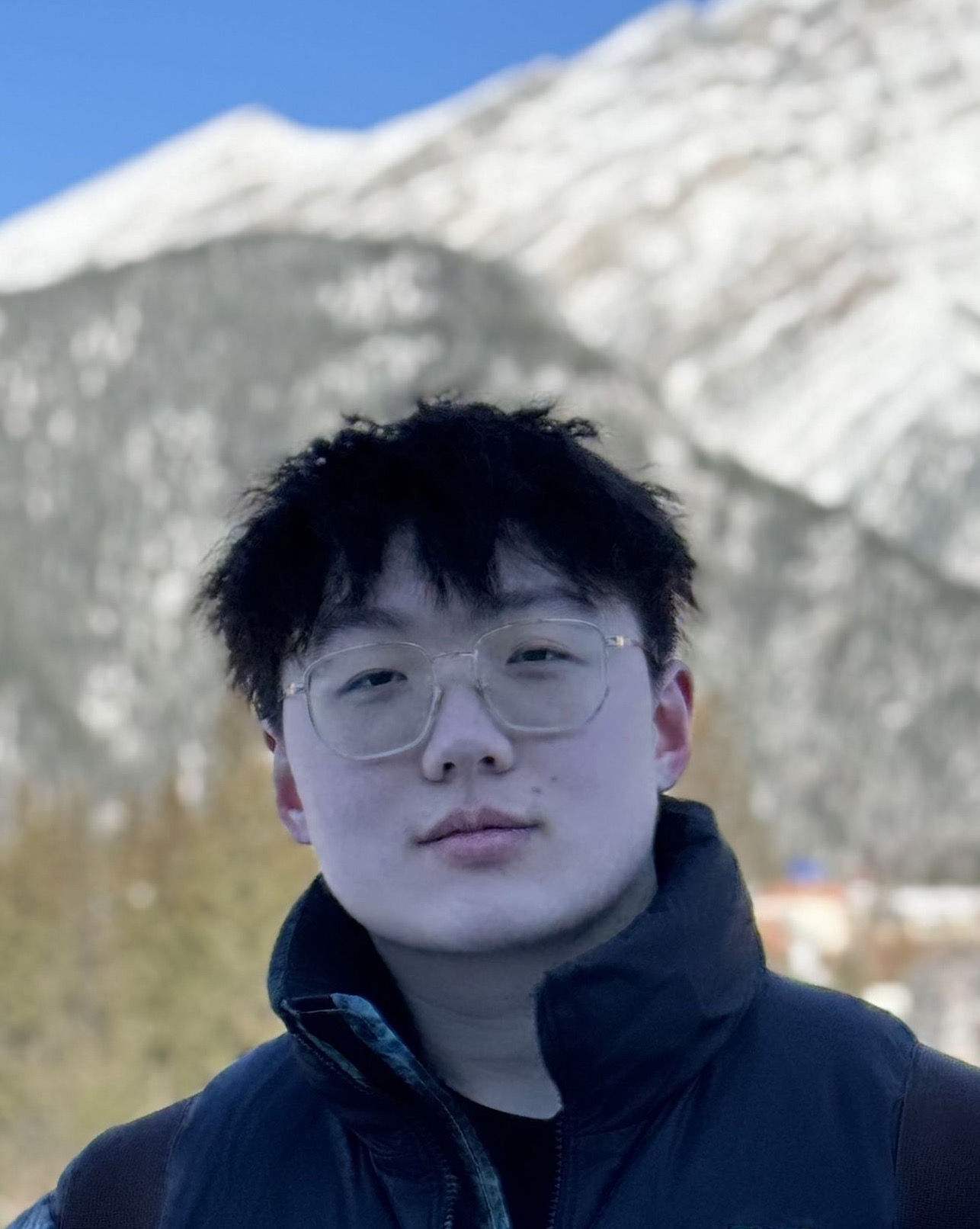}}]{Shengming Zhao} received a B.E. degree from the Beijing University of Posts and Telecommunications, Beijing, China. He is currently pursuing a M.Sc. degree with the University of Alberta, Edmonton, Canada. His research interests include quality assurance of AI systems and automatic code generation.
\end{IEEEbiography}

\vspace{-8mm}

\begin{IEEEbiography}
[{\includegraphics[width=1in,height=1.25in,clip,keepaspectratio]{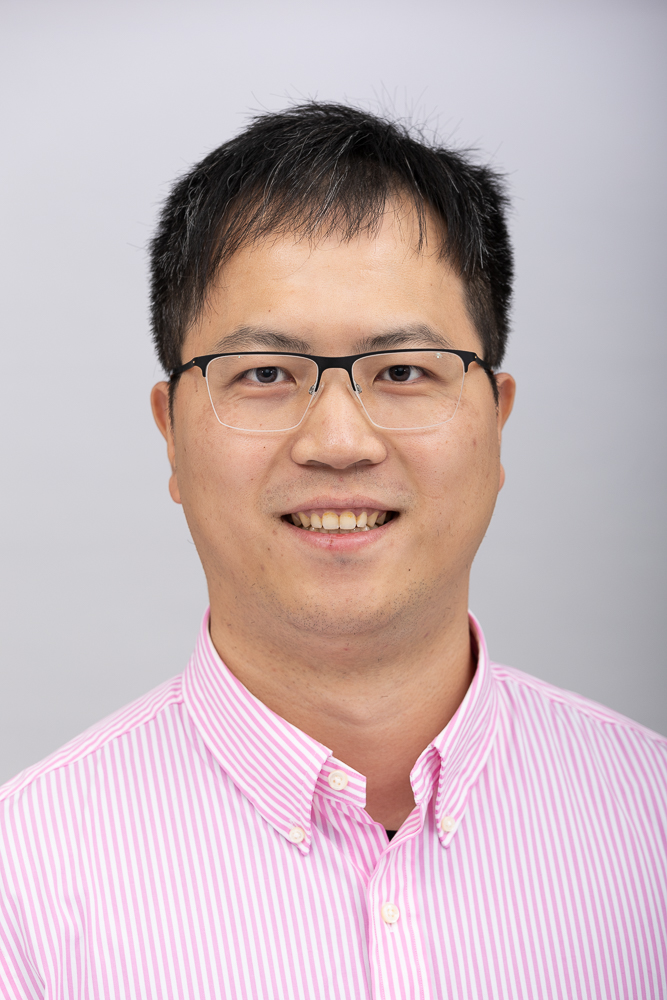}}]{Huaming Chen}(Member, IEEE) received the Ph.D. degree from the University of Wollongong, Wollongong, Australia. He is currently a senior lecturer with the School of Electrical and Computer Engineering, the University of Sydney, Sydney, Australia. His main research interests include software engineering/security, trustworthy AI, and applied machine learning. He regularly serves on the program committees of ACM MM, CCS, ACSAC, SANER, IJCAI, KDD, The Web Conference, SIAM ICDM, ECML/PKDD and so on.
\end{IEEEbiography}

\newpage

\begin{IEEEbiography}[{\includegraphics[width=1in,height=1.25in,clip]{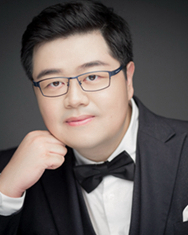}}]{Felix Juefei-Xu} (Member, IEEE) received the Ph.D. degree in Electrical and Computer Engineering from Carnegie Mellon University (CMU), Pittsburgh, PA, USA. Prior to that, he received the M.S. degree in Electrical and Computer Engineering and the M.S. degree in Machine Learning from CMU, and the B.S. degree in Electronic Engineering from Shanghai Jiao Tong University (SJTU), Shanghai, China. Currently, he is a Research Scientist with GenAI at Meta, based in New York City, where he works on robust perception and efficient learning problems in the domain of generative AI. He is also affiliated with New York University as an Adjunct Professor. Previously, he was a Research Scientist with Alibaba Group, based in Sunnyvale, CA. He was the recipient of multiple best or distinguished paper awards, including IJCB 2011, BTAS 2015 and 2016, ASE 2018, and ACCV 2018.
\end{IEEEbiography}

\newpage

\begin{IEEEbiography}[{\includegraphics[width=1in,height=1.25in,clip,keepaspectratio]{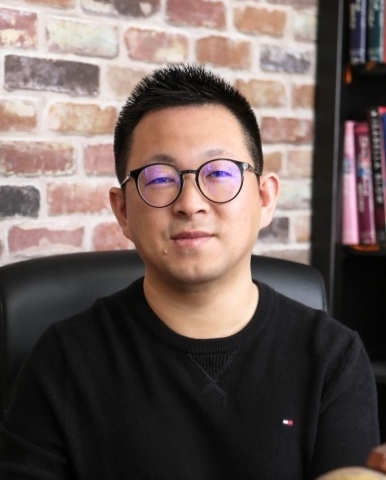}}]{Lei Ma} (Member, IEEE) 
received the B.E. degree from Shanghai Jiao Tong University, Shanghai, China, in 2009, and the M.E. and Ph.D. degrees from The University of Tokyo, Tokyo, Japan, in 2011 and 2014, respectively. He is currently an Associate Professor with The University of Tokyo and the University of Alberta, Edmonton, AB, Canada. He was honorably selected as Canada CIFAR AI Chair and a fellow with Alberta Machine Intelligence Institute (Amii), Edmonton. His research interests include the interdisciplinary fields of software engineering (SE) and trustworthy artificial intelligence, with a special focus on the quality, reliability, safety, and security aspects of AI systems.
\end{IEEEbiography}




\end{document}